\documentclass[aps,prd,11pt,tightenlines,superscriptaddress,nofootinbib,preprintnumbers,notitlepage]{revtex4-1}
 
\pdfoutput=1
\usepackage{amsmath,multirow,graphicx,tabularx,epsfig}
\usepackage{color}
\usepackage{hyperref}

\newcommand\one{\leavevmode\hbox{\small1\normalsize\kern-.33em1}}

\newcommand{\lag}{\mathcal{L}}

\newcommand{\iab}{\text{ab}^{-1}}

\def\slashchar#1{\setbox0=\hbox{$#1$}           
   \dimen0=\wd0                                 
   \setbox1=\hbox{/} \dimen1=\wd1               
   \ifdim\dimen0>\dimen1                        
      \rlap{\hbox to \dimen0{\hfil/\hfil}}      
      #1                                        
   \else                                        
      \rlap{\hbox to \dimen1{\hfil$#1$\hfil}}   
      /                                         
   \fi}

\newcommand{\eg}{\textsl{e.g.}\;}
\newcommand{\ie}{\textsl{i.e.}\;}

\newcommand{\be}{\begin{eqnarray*}}
\newcommand{\ee}{\end{eqnarray*}}

\newcommand{\bee}{\begin{eqnarray}}
\newcommand{\eee}{\end{eqnarray}}
\newcommand{\beeq}{\begin{equation}}
\newcommand{\eeeq}{\end{equation}}

\begin{document}
 
\title{Mad-Maximized Higgs Pair Analyses}

\preprint{FERMILAB-PUB-16-267-T}
\preprint{IPPP/16/72}
\preprint{MCnet-16-28}

\author{Felix Kling}
\affiliation{Department of Physics, University of Arizona, Tucson, USA}
\affiliation{Theoretical Physics Department, Fermilab, Batavia, USA}

\author{Tilman Plehn}
\affiliation{Institut f\"ur Theoretische Physik, Universit\"at Heidelberg, Germany}
 
\author{Peter Schichtel}
\affiliation{Institute for Particle Physics Phenomenology, Durham University, UK}

\begin{abstract}
  We study Higgs pair production with a subsequent decay to a pair of
  photons and a pair of bottoms at the LHC. We use the log-likelihood
  ratio to identify the kinematic regions which either allow us to separate
  the di-Higgs signal from backgrounds or to determine
  the Higgs self-coupling. We find that both regions are separate
  enough to ensure that details of the background modelling will not
  affect the determination of the self-coupling. Assuming dominant
  statistical uncertainties we determine the best precision with which
  the Higgs self-coupling can be probed in this channel. We finally
  comment on the same questions at a future 100 TeV collider.
\end{abstract}

\maketitle

\bigskip 
\bigskip 

\tableofcontents 

\newpage

\section{Introduction}
\label{sec:intro}

\begin{figure}[b!]
  \includegraphics[width=0.65\textwidth]{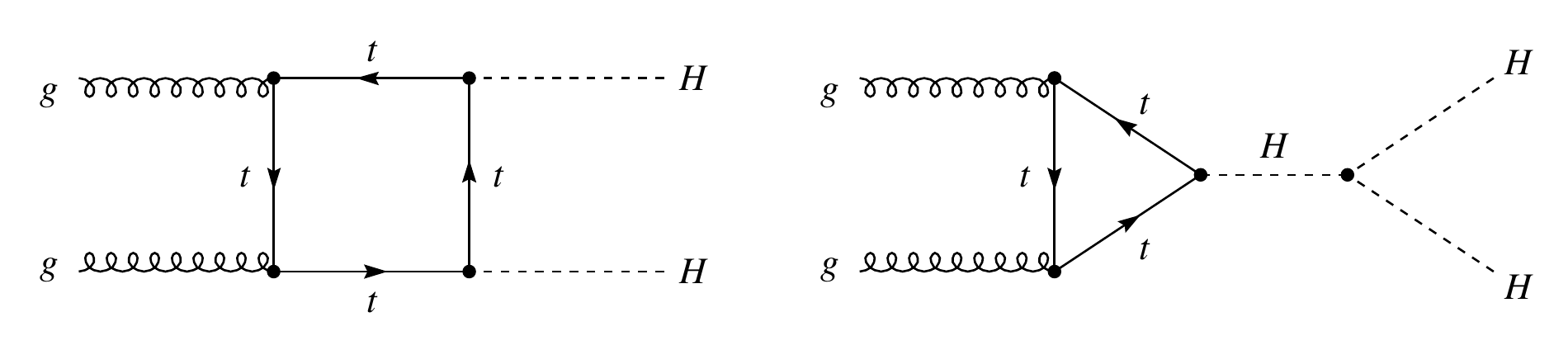}
\caption{Feynman diagrams contributing to Higgs pair production at the
  LHC. Figure from Ref.~\cite{uli1}.}
\label{fig:feynman}
\end{figure}

Higgs pair production is one of the key benchmarks for any new
collider probing the electroweak scale. This includes the LHC towards
large luminosity, but also future electron-positron colliders and
future hadron colliders. At hadron colliders the task is clear: we can
test Higgs pair production in gluon fusion, with continuum
contributions as well as contributions induced by the Higgs
self-coupling. The two Feynman diagrams are shown in
Fig.~\ref{fig:feynman}~\cite{orig,spirix}. In the Standard Model both
diagrams rely on a strong interaction between the Higgs and the top
quark in the loop to give an observable rate for the LHC. This means
that if we want to measure the di-Higgs rate or even the Higgs self
coupling~\cite{uli1,uli2,uli3} from the total rate $\sigma_{gg \to
  HH}$ we need to make an assumption about the top Yukawa
coupling~\cite{review}. For a combined Higgs fit we could for example
assume that the effective gluon-Higgs coupling is only mediated by the
Standard Model quarks, which to date gives a roughly 10\% measurement
of the top Yukawa~\cite{legacy1}.  A model independent precision
measurement of the top Yukawa coupling at the per-cent level will only
be possible at a 100~TeV collider~\cite{nimatron_yt}. Firmly
connecting a possible modification of the Higgs self-coupling to a
modified top Yukawa in a given model is rather hopeless, as can easily
be seen in two-Higgs doublet models. One way to limit the impact of
such an assumption would be to include Higgs pair production as a
probe of the Higgs self-coupling in a global analysis of the Higgs
effective Lagrangian at tree level or at one
loop~\cite{legacy2,hh_d6}.  Another way to at least minimize the
assumption about the top Yukawa is to test kinematic distributions in
Higgs pair production. There are three obvious questions concerning
such an analysis of the Higgs pair kinematics which we tackle in this
paper,
\begin{enumerate}
\item which kinematic features allow us to extract Higgs pair production
  from backgrounds?
\item which kinematic features include information about the Higgs self-coupling?
\item what is the most optimistic LHC sensitivity in the presence of
  QCD backgrounds?
\end{enumerate}
\bigskip

There are two kinematic regimes which are well known to carry
information on the Higgs self-coupling.  Both of them exploit the
(largely) destructive interference between the two graphs shown in
Fig.~\ref{fig:feynman}. As we will illustrate later, over most of
phase space the continuum contribution dominates. One phase space
region where the triangle diagram become comparable is close to
threshold~\cite{spirix,uli1}; if we denote the Higgs effective
Higgs--gluon Lagrangian in terms of the gluon field strength $G_{\mu
  \nu}$ as~\cite{low_energy}
\begin{align}
\lag_{ggH} 
=\frac{\alpha_s}{12\pi} \; G^{\mu\nu}G_{\mu\nu} \;
 \log \left(1+\frac{H}{v} \right) 
= G^{\mu\nu}G_{\mu\nu} \; \frac{\alpha_s}{12 \pi v} \;
\left( H - \frac{H^2}{2v} + \ldots \right) \; ,
\label{eq:higgs_eff}
\end{align}
we can write the amplitude for Higgs pair production as
\begin{align}
\mathcal{A} \propto 
\frac{\alpha_s}{12 \pi v} \; 
\left( \frac{\lambda}{s-m_H^2} - \frac{1}{v} \right) 
\longrightarrow
\frac{\alpha_s}{12 \pi v^2} \; 
\left( \frac{3m_H^2}{3m_H^2} -1 \right) = 0 
\qquad \text{for} \quad 
m_{HH} \to 2 m_H \; .
\label{eq:higgs_pair}
\end{align}
The exact cancellation is linked to the Standard Model value of the
physical Higgs self-coupling $\lambda = 3m_H^2/v$. Note that while the
heavy top approximation is well known for giving completely wrong kinematic
distributions for Higgs pair production~\cite{uli1}, it correctly
predicts the threshold region. If we rely on this distribution, the
strategy behind an LHC analysis will be to rule out large deviations
from the Standard Model Higgs self-coupling based on the fact that any
such deviation leads to a strongly enhanced production cross section.

The second sensitive kinematic regime where the two contributions
shown in Fig.~\ref{fig:feynman} become comparable is boosted Higgs
pair production~\cite{boosted}. Top threshold corrections to the
triangle diagram are strongly enhanced around $m_{HH} = 2 m_t$. We can
translate this into a condition for the transverse momentum of the
SM-like Higgs, where we find that around $p_{T,H} \sim 100$~GeV the
combined process develops a minimum for large Higgs
self-couplings. The fact that there exist two relatively un-correlated
useful kinematic distributions is of course not surprising for an
effectively 2-body final state.\bigskip

At the LHC, we can go through all Higgs decay channels and test their
combinations as possible di-Higgs signatures. The most promising
channel for a SM-like pair of Higgs bosons is most likely the
$b\bar{b}\gamma\gamma$ final state~\cite{uli3,vernon,atlas,cms}. Its
great advantage is that we can easily reconstruct one of the two Higgs
bosons and that the QCD continuum background can be measured in
control regions. In addition, it should be possible to use the
$b\bar{b} \tau \tau$ final state~\cite{uli2,boosted}, if tau-tagging
will show a sufficient performance. The combination
$b\bar{b}WW$~\cite{bbww} will only work if we can suppress the
$t\bar{t}$ background, while the $4b$~\cite{uli2,bbbb} and the
original $4W$~\cite{uli1,wwww} signatures are unlikely to contribute
significantly for a pair of SM-like Higgs bosons. Finally, the
$b\bar{b} \mu \mu$ shares many beneficial features with the $b\bar{b}
\gamma \gamma$ channel~\cite{uli3}, but will be further suppressed by
the muon branching ratio. The picture changes if we consider either
resonant Higgs pair production~\cite{uli3,resonant} or strongly
interacting Higgs pair production~\cite{strong}. For a pair of SM-like
Higgs bosons there exists a large number of theoretical precision
calculations~\cite{review}, including NLO~\cite{nlo} and
NNLO~\cite{nnlo} predictions for the differential rates. Those will be
crucial if we want to study this production process in spite of the
top-Yukawa-infected total rate prediction.\bigskip

In this paper we first generalize \textsc{MadMax} from computing
maximum signal significances, globally or distributed over phase
space, to comparing two general hypotheses in
Sec.~\ref{sec:intro_stats} and~\ref{sec:intro_fft}. In
Sec.~\ref{sec:sb} we then ask the question which kinematic information
should allow us to extract Higgs pair production from the QCD
backgrounds. We also recapitulate how some of the key features arise
from the combination of the triangle and box diagrams shown in
Fig.~\ref{fig:feynman}. Next, we test which kinematic distributions
allow us to measure the Higgs self-coupling at the LHC in
Sec.~\ref{sec:self}. In Appendix~\ref{app:100} we provide the
corresponding information for a future 100~TeV collider.

\subsection{MadMax}
\label{sec:intro_stats}

The \textsc{MadMax}~\cite{madmax1,madmax2} approach of calculating
significance distributions for kinematic observables is based on the
Neyman--Pearson lemma: the likelihood ratio is the most powerful test
statistic for a hypothesis test between a simple null hypothesis ---
for example background only --- and an alternate hypothesis --- for
example signal plus background~\cite{proof}. Maximum power is formally
defined as the minimum probability for false negative error for a
given probability of false positive.

We have established that we can define and compute the maximum
significance of a signal-plus-background process as compared to the
background-only hypothesis using the standard Monte Carlo tools. As an
example we studied Higgs decays to muons in weak boson
fusion~\cite{madmax1}. Our results can be taken as a benchmark for the
performance of multi-variate analysis techniques at the LHC, including
the matrix element method. In a second step we used the same method to
determine which phase space regions contribute to this maximum
significance, for example in boosted Higgs production in the $ZH$ or
$t\bar{t}H$ channels~\cite{madmax2}. Such a study allows us to
determine how much of the distinguishing power of a multi-variate
analysis comes from phase space regions which are systematically and
theoretically under control. In this paper we extend this
approach to test two signal hypotheses, with the technical
complication that over phase space the expected ratio of events can
lie on either side of unity.\bigskip

In general, the likelihood of observing $n$ events assuming a
hypothesis $H_0$ is given by the Poisson distribution
$\text{Pois}(n|n_0)=e^{-n_0} \, n_0^n/n!$. We can generalize this
counting experiment by introducing a observable $x$, where we assume
that $H_0$ is described by the normalized distribution
$f_0(x)$. Similarly, the alternative hypothesis $H_1$ is described by
$f_1(x)$. Each likelihood can be factorized into the Poisson
likelihood to observe an event and the normalized event likelihood
$f_{0,1}(x)$. In combination they give the single-event log-likelihood
ratio (LLR)
\begin{align}
q_{n=1}(x) &= \log \; \frac{L(x|H_1)}{L(x|H_0)}
      = \log \; \frac{\text{Pois}(1|n_1)    \; f_1(x)}
                     {\text{Pois}(1|n_0)  \; f_0(x)}  \notag \\
     &= (n_0 - n_1)  
        +  \log \frac{n_1}{n_0} 
        +  \log \frac{f_1(x)}{f_0(x)} \; .
\label{eq:llr1}
\end{align}
If we know $f_0(x)$ and $f_1(x)$ from Monte Carlo simulations, we can
compute the above LLR. A very simple structure containing only the last
term appears, when we only rely on kinematic distributions and not on
the total rate. We can generalize Eq.\eqref{eq:llr1} to the likelihood
of observing $n$ events in a phase space configuration $\vec
x=\{x_j\}$. The normalized event likelihood is now a product of $n$
likelihoods at the corresponding configurations, so we find
\begin{align}
q_n(\vec x) &= \log \; \frac{L(\vec x|H_1)}{L(\vec x|H_0)}
      = \log \; \frac{\text{Pois}(n|n_1)    \; \prod_{j=1}^n f_1(x_j)}
                     {\text{Pois}(n|n_0)  \; \prod_{j=1}^n f_0(x_j)}  \notag \\
     &= (n_0 - n_1)  +  \sum_{j=1}^{n} \log \frac{n_1 f_1(x_j)}{n_0 f_0(x_j)} \; .
\label{eq:llrn}
\end{align}
The combined LLR is additive when we include more than one event. If
the argument of the logarithm is allowed to be greater as well as
smaller as unity, depending on the position in phase space, their
contribution to the over-all LLR may cancel.

If we want to use the LLR to distinguish two hypotheses we need to
evaluate our events as a function of the LLR, given either $H_0$ or $H_1$.
Assuming for example the hypothesis $H_0$ we can integrate over the
entire phase space $\vec x$ with the normalized event weight
$d\sigma_0(\vec x)/\sigma_{0, \text{tot}}$ and generate a LLR distribution
based on the relation $q_n(\vec x)$ given in Eq.\eqref{eq:llr1},
\begin{align}
\rho_{0,n=1}(q) 
= \int dx \; f_0(x) \;
             \delta( q_{n=1}(x) - q) \; .
\label{eq:rho_single}
\end{align}
The corresponding likelihood distributions for $n$ events and combined
for all possible outcomes $n$ are given by a convolution in $q$-space
\begin{align}
  \rho_{0,n}(q) &= \rho_{0,n=1} \otimes \rho_{0,n=1} \otimes \cdots \otimes
  \rho_{0,n=1} \notag \\ 
  \rho_0(q) &= \sum_n \text{Pois}(n|n_0) \;
  \rho_{0,n}(q) \; .
\label{eq:convolution}
\end{align} 
The numerical evaluation of such a convolution is best done in Fourier
space and will be the topic of the next section.

To compute significance distributions as a function of any phase space
variable we use the same procedure as for the maximum
significance. However, we restrict the events we use for the
construction of $\rho_{0,n=1}(q)$ to be those which populate the phase
space of a given bin of an observable of interest. We then iterate for
each bin of that observable to fill the differential significance
histogram.

\subsection{Computing likelihood distributions}
\label{sec:intro_fft}

To compute the likelihood distribution we rely on a set of simulated
events covering the entire phase space for each of the hypotheses
$H_0$ and $H_1$. Following Eq.\eqref{eq:rho_single} we first construct
$\rho_{0,n=1}(q)$ and $\rho_{1,n=1}(q)$. The convolution in LLR space
can best be evaluated through a Fourier transform with $q \to \bar{q}$
and $\rho_{0,n} \to \bar{\rho}_{0,n}$. The original convolution in
Eq.\eqref{eq:convolution} turns into a product, namely
\begin{align}
\bar{\rho}_0 
= \sum_n \text{Pois}(n|n_0) \; \bar{\rho}_{0,n}
= e^{-n_0} \, \sum_n \frac{n_0^n}{n!} \; \bar{\rho}_{0,n=1}^n
= e^{n_0 \left( \bar{\rho}_{0,n=1} -1 \right)} \;,
\end{align}
where we have chosen, here and in the following, the $H_0$ hypothesis
as a representative. The logic for $H_1$ follows in analogy. We then
need to transform these distributions back into $q$ space and
numerically compute confidence levels by integrating over the relevant
$q$ range.\bigskip

Following the structure of Eq.\eqref{eq:convolution} we know that fast
Fourier transforms based on the discrete Fourier transform (DFT)
should be a convenient tool to numerically generate the full event
likelihood distribution $\rho_0(q)$. DFTs work on an array of numbers
$a_j$ where $j\in[0,N-1]$ for fixed $N$. For a discretized function
the Fourier transformation turns into
\begin{align}
  \bar{a}_k &= \sum \limits^{N-1}_{j=0} a_j \, \text{exp} \left[
    -2\pi\frac{i\,j\,k}{N} \right]\,,
\end{align}
where $\bar{a}_k$ denotes the discrete Fourier transform on the same
size of array $k\in[0,N-1]$~\cite{spectralmethods}. There
are some points which we need to take care of if we are to use this
formalism. To use a DFT we need to project our function $\rho_{0,n=1}$
into a binned histogram $\rho_{0,n=1,\text{binned}}$, where in
practice we use up to 400 bins for the projection. An issue with the
full event likelihood distribution is that it will not have the same
support in LLR space as the single event likelihood. We need to
pre-define an interval in LLR space to perform the complete
computation on the same array. Using simple Poisson counting we can
estimate the length of this interval to be~\cite{lepstats}.
\begin{align}
  q_\text{length} &= \sum_{(q,f_0)  \in \rho_{0,n=1,\text{binned}}} 
  \left|q\right| \left( \,
  n_0 f_0 + \sigma^\text{max}
  \sqrt{n_0 f_0} \right).
\label{eq:qmax}
\end{align}
The additional factor in parentheses encodes the feature that a
Poisson distribution is most likely localized around the expectation
value $n_0$ with a standard deviation of $\sqrt{n_0}$. For each step
or convolution a given $q$-value will typically move by a factor
$n_0 + \sigma^\text{max}\sqrt{n_0}$, where $\sigma^\text{max}$ counts
the number of maximally expected standard deviations. By summing over
all bins of $\rho_{0,n=1,\text{binned}}$ and using the absolute value
of $q$, we ensure that the full likelihood distribution will fit into
the fixed size array. To be on the conservative side we use
$\sigma^\text{max}=8$ and multiply the final result by a factor of
two. Because both, $\rho_0$ and $\rho_1$ have to be mapped to the same
array to allow for a meaningful computation of confidence levels we
also need to replace $n_0 f_0$ in Eq.\eqref{eq:qmax} by its maximum
values for the two hypotheses.  This length we divide into $2^{19}$
bins $a_j$, see also Ref.~\cite{lepstats} for more details.

The main difference between this analysis and the original
signal-background study of Ref.~\cite{madmax1,madmax2} is that the LLR
$q$ can switch signs.  As we do not restrict the allowed values of $q$
when we compute the single event likelihood distribution from MC, we
also need to know how to map $\rho_{0,n=1,\text{binned}}$ onto the
array $a_j$. By construction, the first bin in the DFT array
corresponds to zero on the
$q$-axis~\cite{spectralmethods}. The DFT represents a
Fourier transform on a finite interval, which means periodic boundary
conditions. If we are to use an arbitrary single event likelihood
distribution where negative values of $q$ are allowed, we need to
respect these boundary conditions by moving all negative values to the
very right of the $a_j$ array, while the space in between will be
filled with zeros. This way we can make sure to compute the correct
full event likelihood distribution. To do so we use the standard fast
Fourier transforms as implemented, for example, in the Python package
\textsc{Scipy}.

\subsection{Analysis setup}
\label{sec:intro_ana}

\begin{figure}[b!]
  \includegraphics[width=0.30\textwidth]{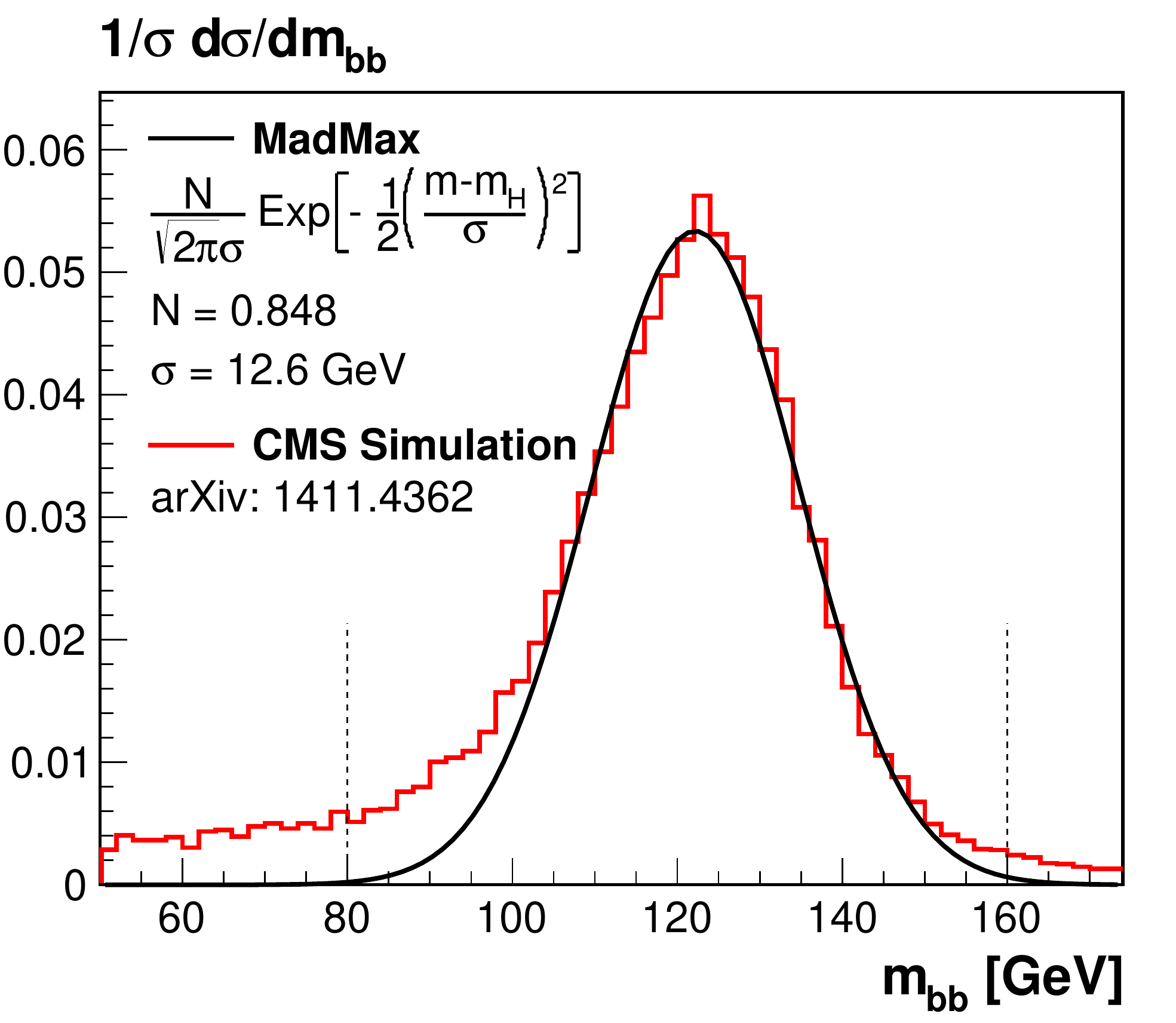}
  \hspace*{0.02\textwidth}
  \includegraphics[width=0.30\textwidth]{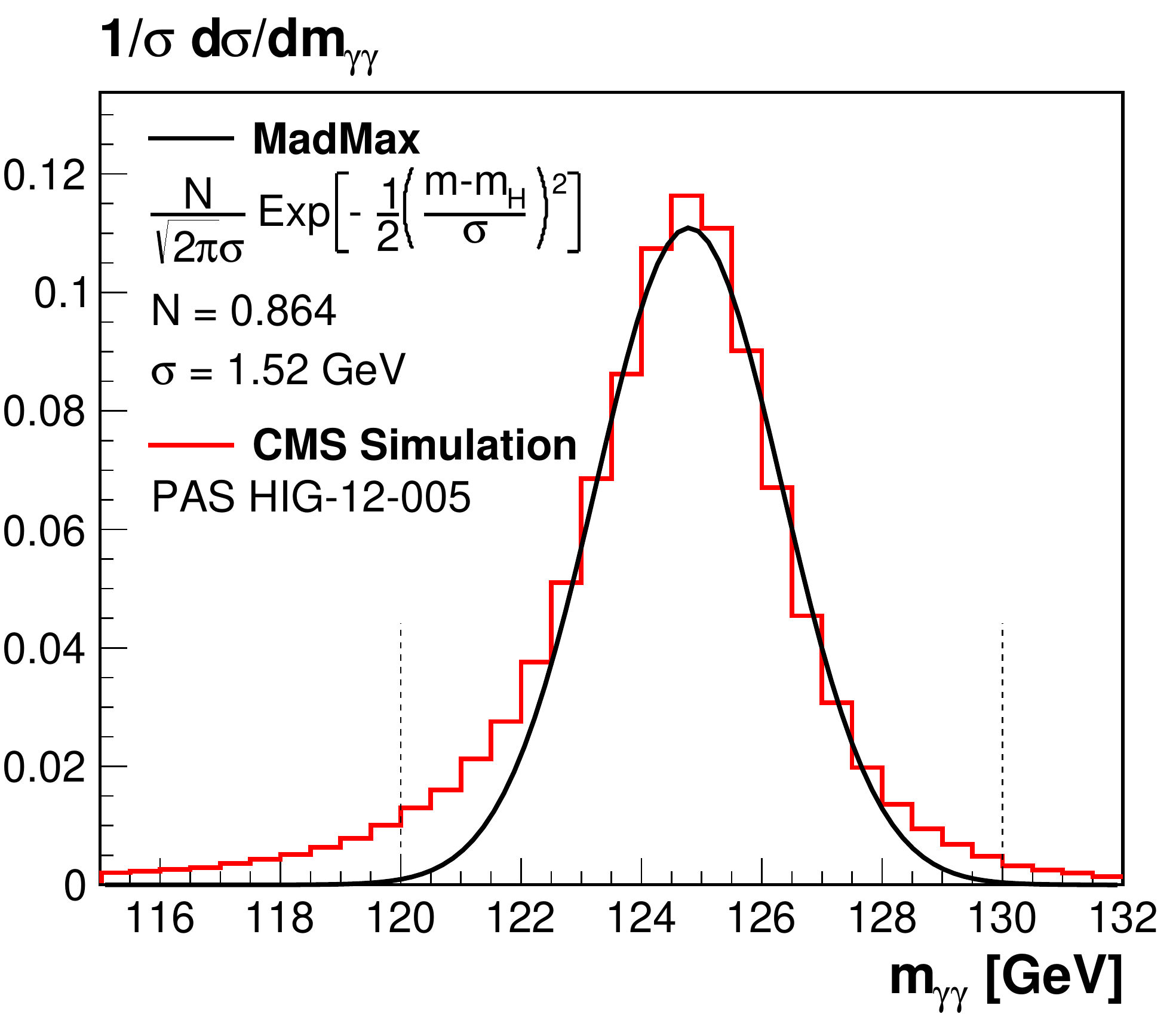}
  \hspace*{0.02\textwidth}
  \includegraphics[width=0.30\textwidth]{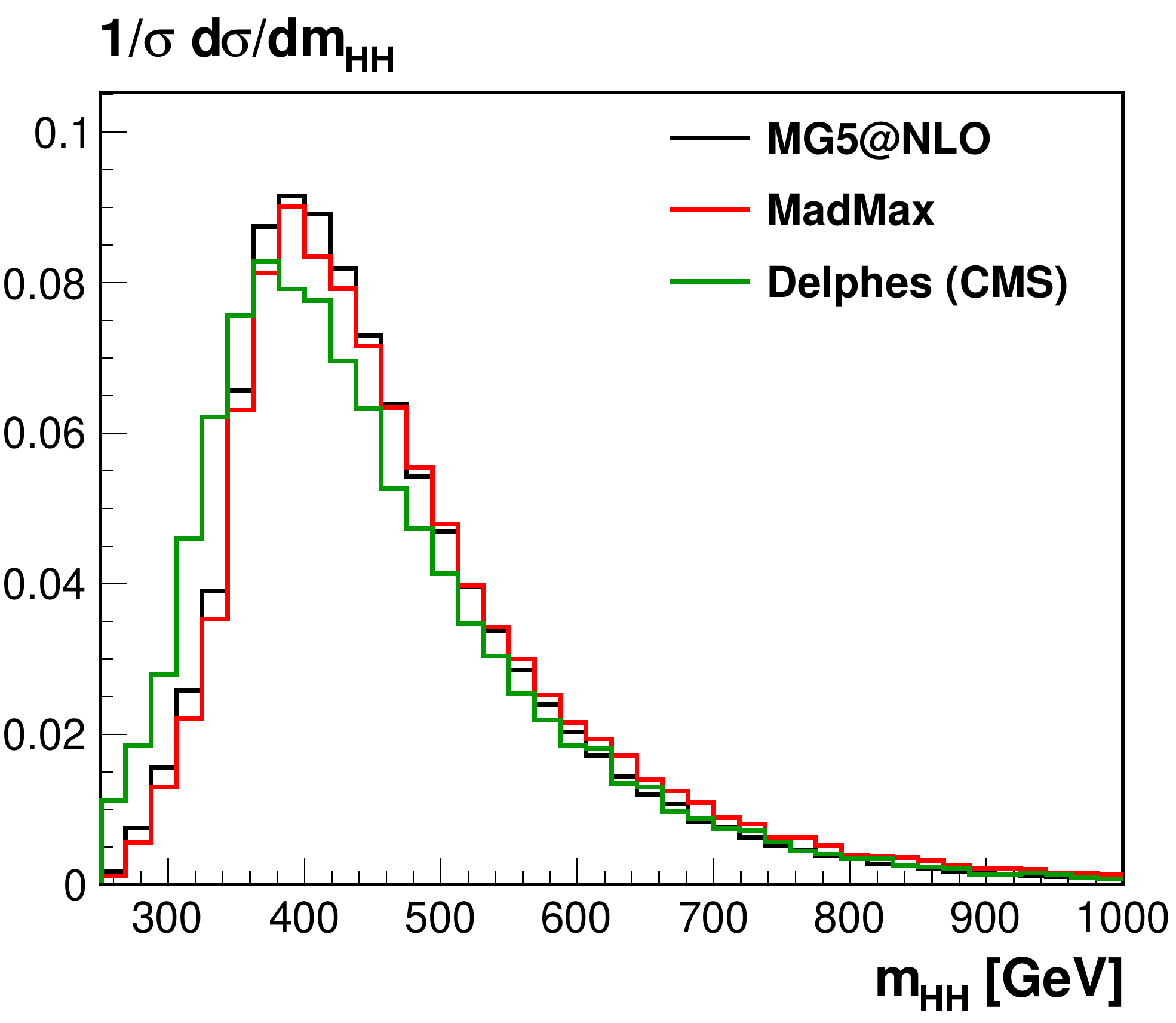}
  \includegraphics[width=0.30\textwidth]{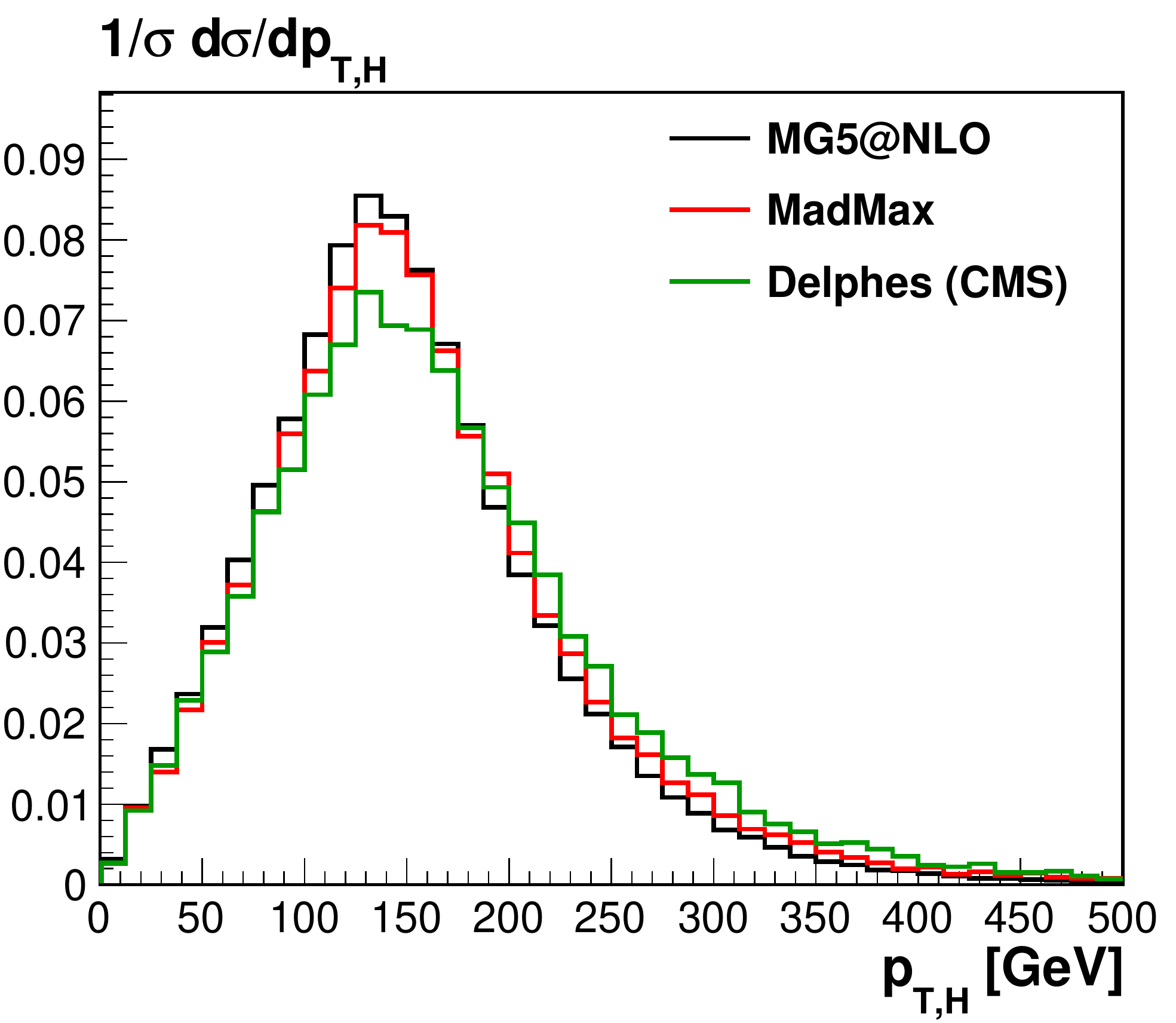}
  \hspace*{0.02\textwidth}
  \includegraphics[width=0.30\textwidth]{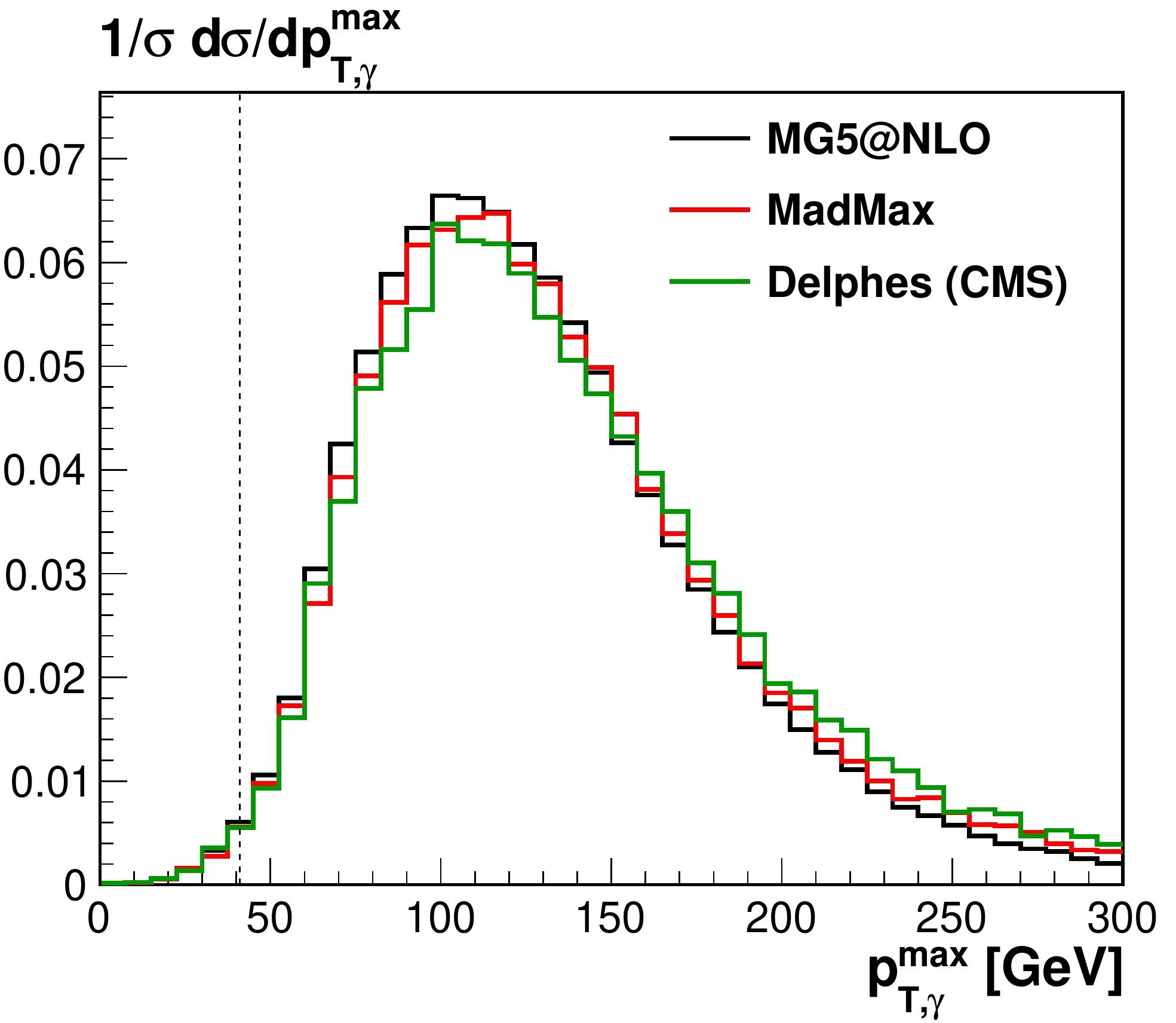}
  \hspace*{0.02\textwidth}
  \includegraphics[width=0.30\textwidth]{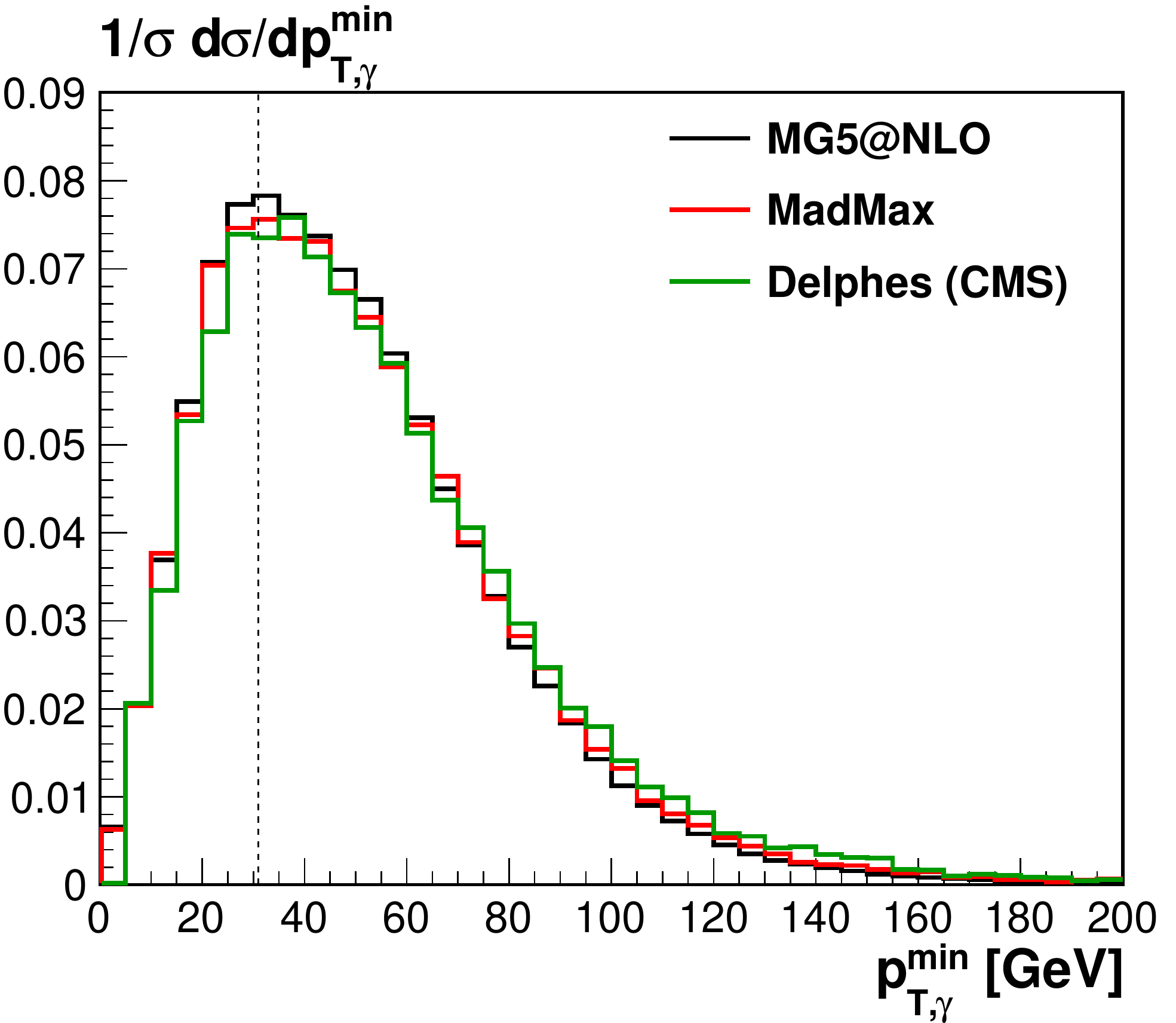}
  \caption{First two panels: normalized invariant mass distributions
    of the di-jet and di-photon systems as used in \textsc{MadMax} and
    from CMS~\cite{Vernieri:2014wfa,CMS:2016zjv}. Next four panels:
    di-Higgs invariant mass, transverse momentum spectra used in
    \textsc{MadMax}, including NLO corrections~\cite{hh_madgraph}, and
    including the fast detector simulation
    \textsc{Delphes}~\cite{delphes}.}
\label{fig:smear}
\end{figure}

While our analysis closely follows the original proposal in
Ref.~\cite{uli3}, we ensure that in particular our detector simulation
corresponds to the current state of the art. For the signal we use the
NLO production cross section $\sigma(pp\to HH) = 34.8$~fb at 14~TeV
center of mass energy~\cite{hh_madgraph}.  For the particular
$b\bar{b}\gamma\gamma$ final state the SM rate prediction is
$0.092$~fb. Unlike in the CMS study of Ref.~\cite{cms} we only
consider the high-purity category where both $b$-jets are tagged and
which should dominate the significance at the high-luminosity LHC.
Following Ref.~\cite{uli3} and the CMS study~\cite{cms} main
backgrounds are the non-resonant $\gamma\gamma+2$~jets process ---
including $\gamma\gamma c \bar{c}$ --- and $\gamma j +2$~jets
production.  Throughout our analysis we will see that these QCD
continuum backgrounds have essentially the same shape, while their
relative size depends on the photon and bottom identification. This
also means that adding mis-tagged backgrounds can easily be taken into
account by scaling the total QCD background rate.

Of the different Higgs processes we include the irreducible, resonant $ZH \to
bb\gamma\gamma$ background, because it can be hard
to separate from the background. Assuming $\sigma(ZH)=0.9861$~pb,
BR$(Z\to bb)=0.1512$ and BR$(H\to\gamma\gamma)=0.00227$, its cross
section is $\sigma(ZH\to bb\gamma\gamma)=3.38$~fb. The $Z$-peak is
modeled by a double gaussian fit to the CMS study~\cite{Zpeak}.
All remaining non-resonant backgrounds as well as resonant backgrounds
containing $H\to \gamma\gamma$ we assume to be negligible or
comparably easy to control. This also applies to the reducible
$t\bar{t} \gamma \gamma$ background, for which we would need to
update the \textsc{MadMax} analysis to model the crucial QCD jet
activity in the signal and background events.

For the two $b$-quarks and two photons in the final state we need to
simulate the detector performance.  We use a $p_T$-dependent and
$\eta$-dependent $b$-tagging rate with $\epsilon_b \approx 0.7$ at
high $p_T>100$~GeV in the barrel, respectively.  Towards larger $p_T$
the $b$-tagging efficiency will be reduced.  For our choice of
efficiencies the $jj\gamma\gamma$ background ($j=u,d,s,g$) is
negligible. If instead we choose a constant $b$-tagging efficiency and
a $p_T$-dependent misidentification rate the light-flavor backgrounds
in the low-$p_T$ regime increase significantly.  Similarly, the photon
identification efficiency and fake rate due to jets depends on the
barrel versus end-cap position, the transverse momentum, and the
parton forming the jet.  All tagging efficiencies used in this
analysis are described in Appendix~\ref{app:sim}.\bigskip

To simulate the detector effects in \textsc{MadMax} we modify
\textsc{MadGraph5}~\cite{madgraph5} to smear the Higgs propagator like
the square root of a Gaussian distribution. This allows us to
reproduce the measured smearing of the CMS detector for
$b\bar{b}$~\cite{Vernieri:2014wfa} and $\gamma \gamma$
pairs~\cite{CMS:2016zjv}.  In the upper panels of Fig.~\ref{fig:smear}
we show that our prescription is adequate for the peak region, but
faces limitations for the tails of the invariant mass
distributions. However, these tails will hardly contribute to the
signal significance. All we need to do is account for the loss of
signal rate through these tails, $84.8\%$ for the $b\bar{b}$ peak and
$86.4\%$ for the $\gamma\gamma$ peak.

Our choice of trigger cuts is motivated by CMS~\cite{cms}:
$p_{T,\gamma_1}> m_{\gamma\gamma}/3 \approx 41$~GeV, $p_{T,\gamma_2}>
m_{\gamma\gamma}/4 \approx 31$~GeV, $|\eta_\gamma|<2.5$, $\Delta
R_{\gamma\gamma,\gamma j,jj}>0.4$, $m_{\gamma\gamma} =
100~...~180$~GeV, $p_{T,j}>25$~GeV, $|\eta_j|<2.4$, and
$m_{jj}=60~...~180$~GeV. Our invariant mass windows
$m_{bb}=80~...~160$~GeV and $m_{\gamma\gamma}=120~...~130$~GeV are
designed to fully contain the peaks, as illustrated in the first two
panels of Fig.~\ref{fig:smear}. We apply these mass windows for the
continuum background simulation throughout or analysis, which means
that our background kinematics will not fully reproduce the QCD
features.

Also in Fig.~\ref{fig:smear} we show the $m_{HH}$ distribution and the
different transverse momentum spectra entering our analysis.  The
\textsc{MadMax} model is based on a a loop-improved
approach~\cite{hh_madgraph} which includes the NLO form factors
presented in Ref.~\cite{spirix}.  This model is also used by the ATLAS
collaboration~\cite{atlas}.  One reference curve shows the full loop
calculation using \textsc{MG5-aMC\@@NLO}~\cite{nlo_madgraph}, another a fast
detector simulation with \textsc{Delphes}~\cite{delphes}.  Even though
our simulation might not correspond to a precision prediction at the
per-cent level, we see that it reproduces the next-to-leading order
results well.

\section{Standard Model signal vs. background}
\label{sec:sb}

The first question we want to address in this study is: \textsl{which
kinematic features allow us to extract Higgs pair production from
backgrounds?} In this section we identify the Higgs self-coupling with
its Standard Model value $\lambda_\text{SM} = 3m_H^2/v$, ensuring the
perfect threshold cancellation in the heavy top limit shown in
Eq.\eqref{eq:higgs_pair}. In our statistics language 
the null-hypothesis $H_0$ is continuum backgrounds only, while the
hypothesis $H_1$ is defined by Standard Model signal plus backgrounds. 

\begin{figure}[t]
  \includegraphics[width=0.3\textwidth]{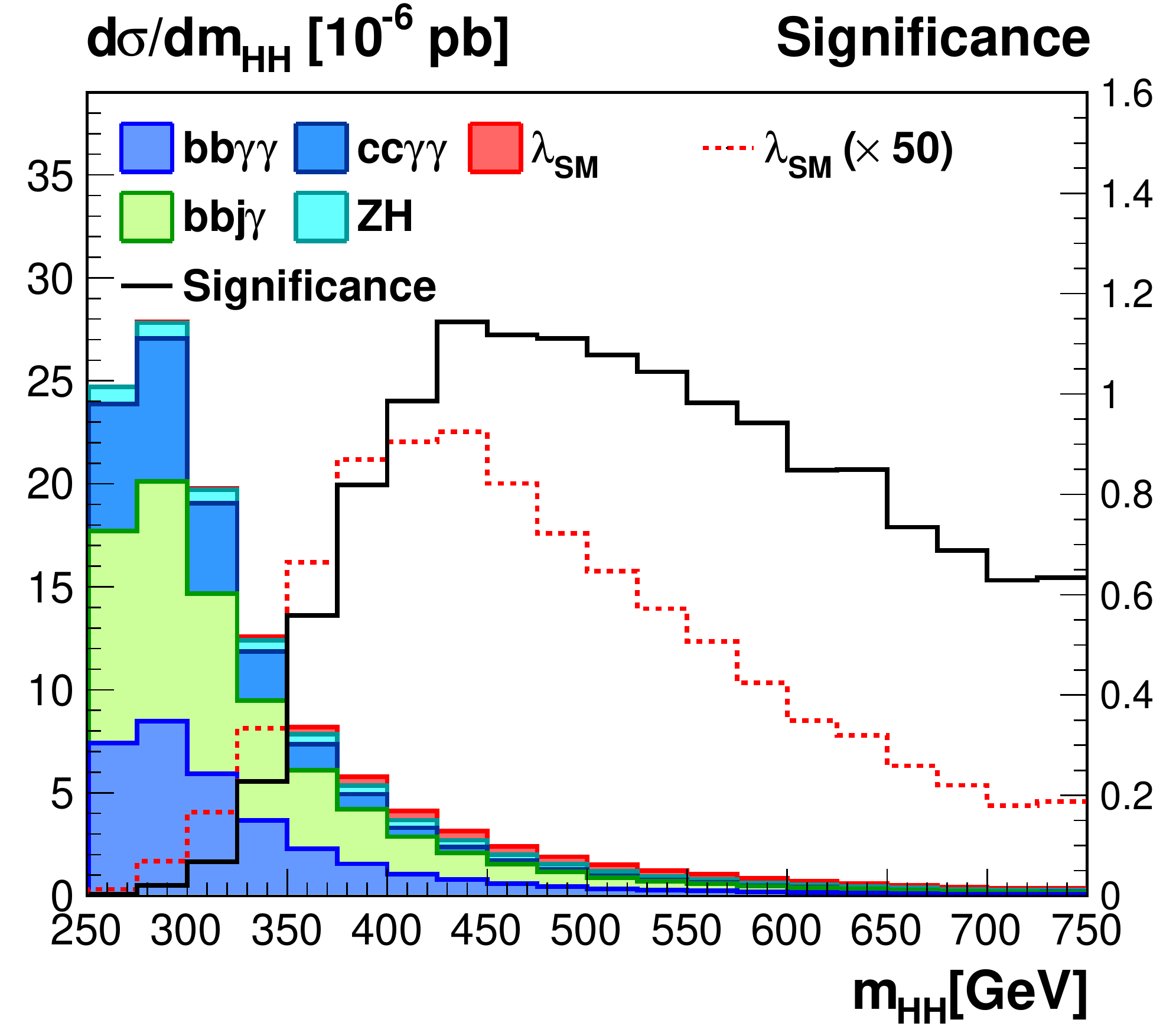}
  \hspace*{0.02\textwidth}
  \includegraphics[width=0.3\textwidth]{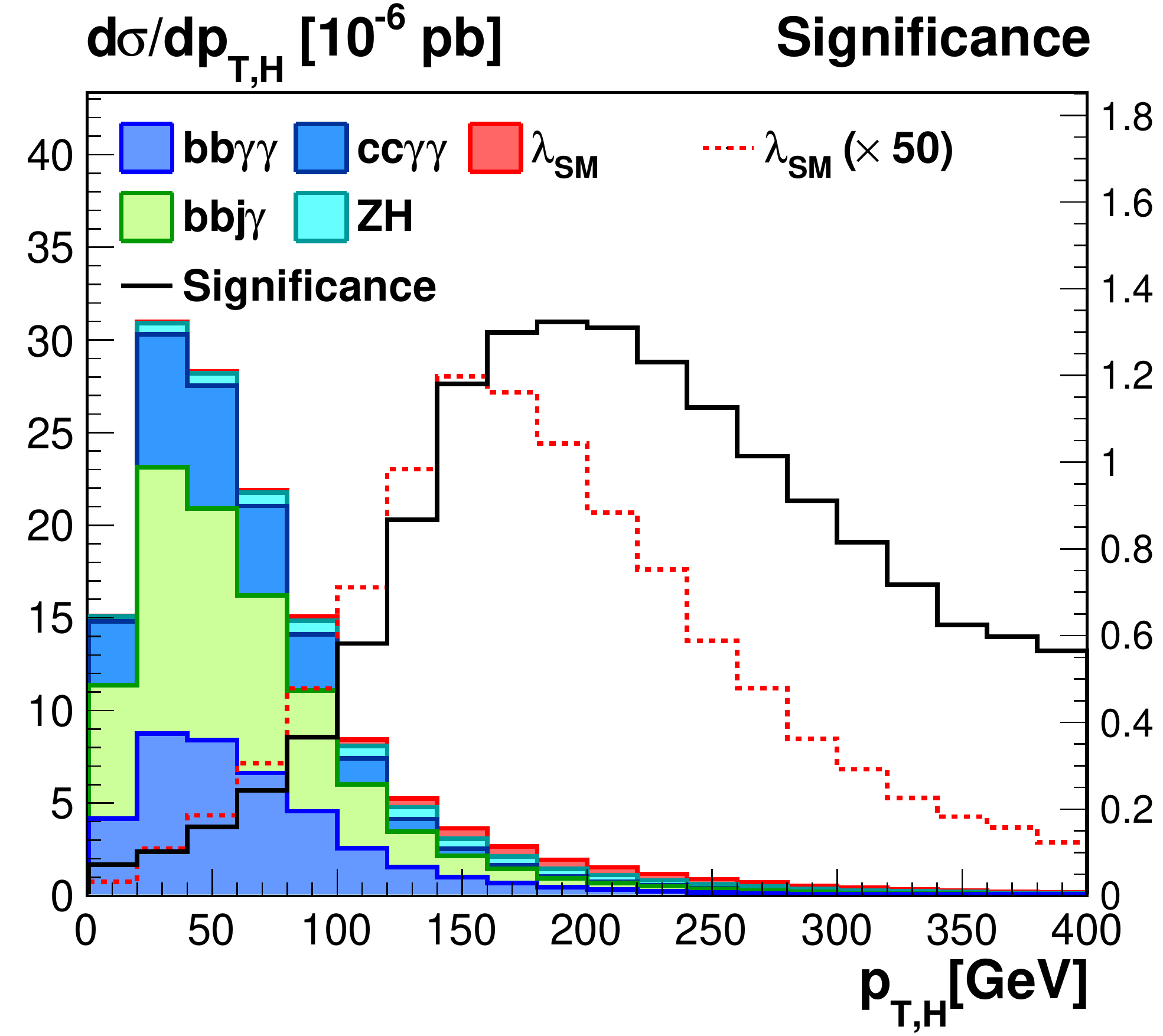}
  \hspace*{0.02\textwidth}
  \includegraphics[width=0.3\textwidth]{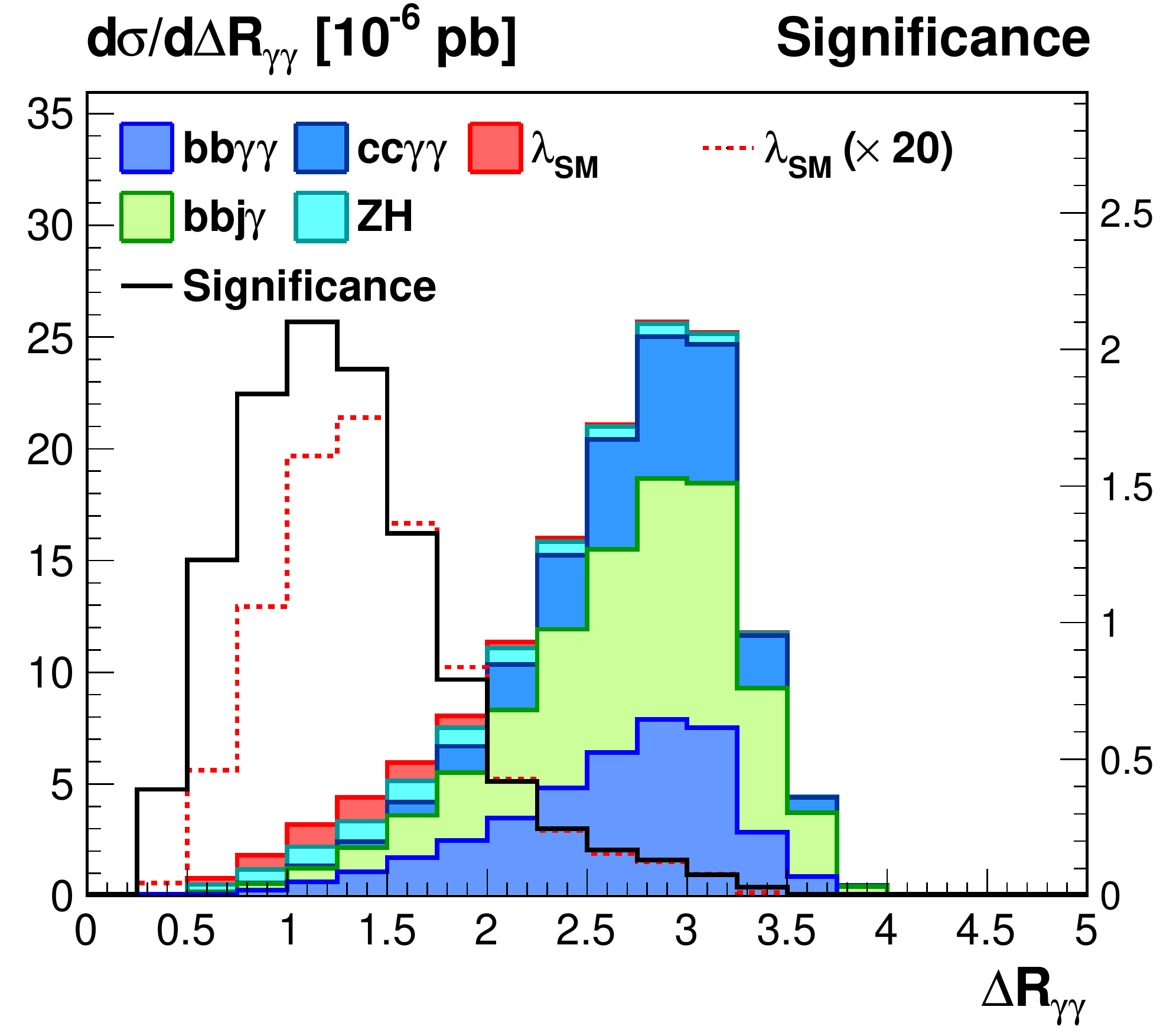}
  \includegraphics[width=0.3\textwidth]{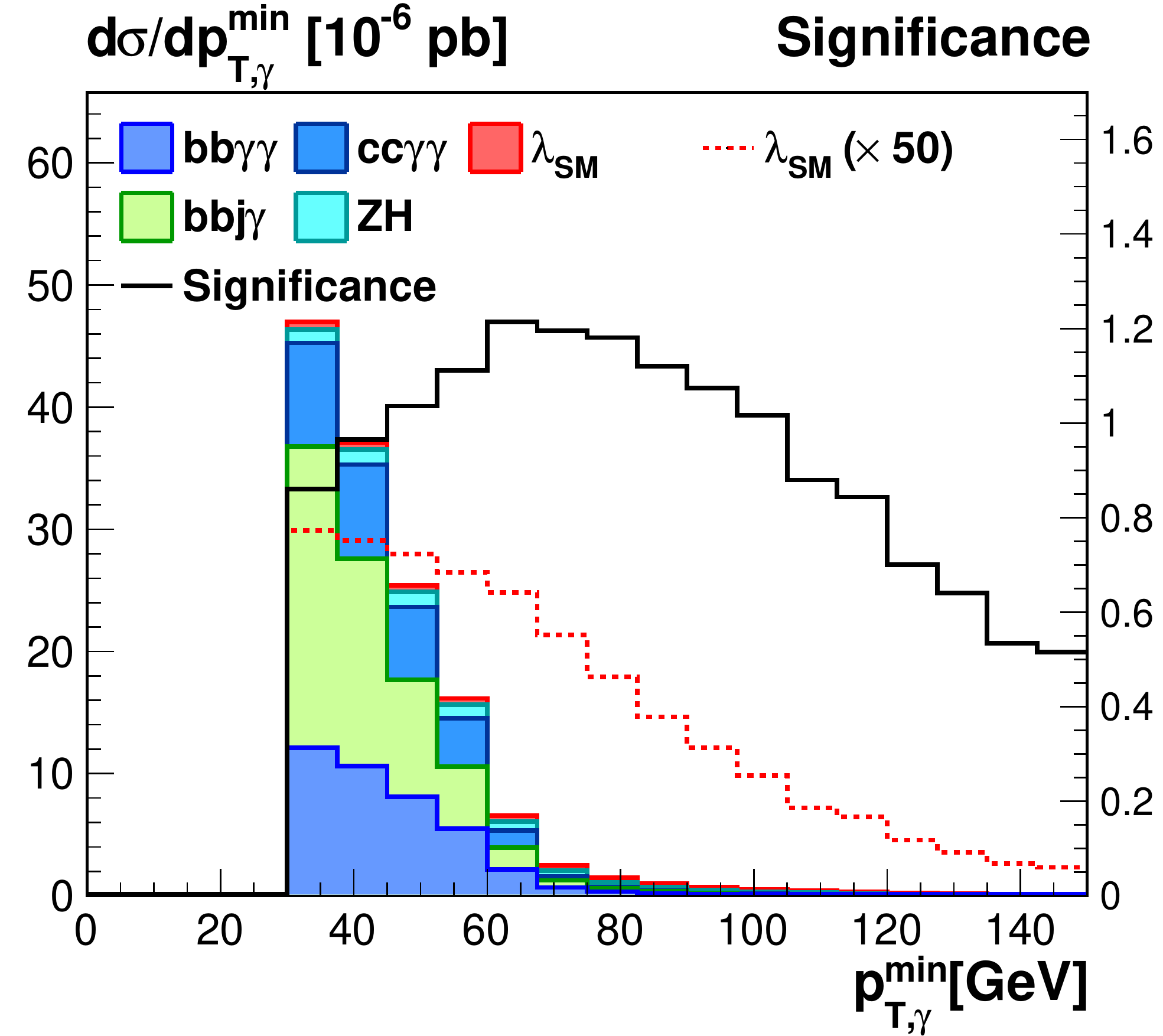}
  \hspace*{0.02\textwidth}
  \includegraphics[width=0.3\textwidth]{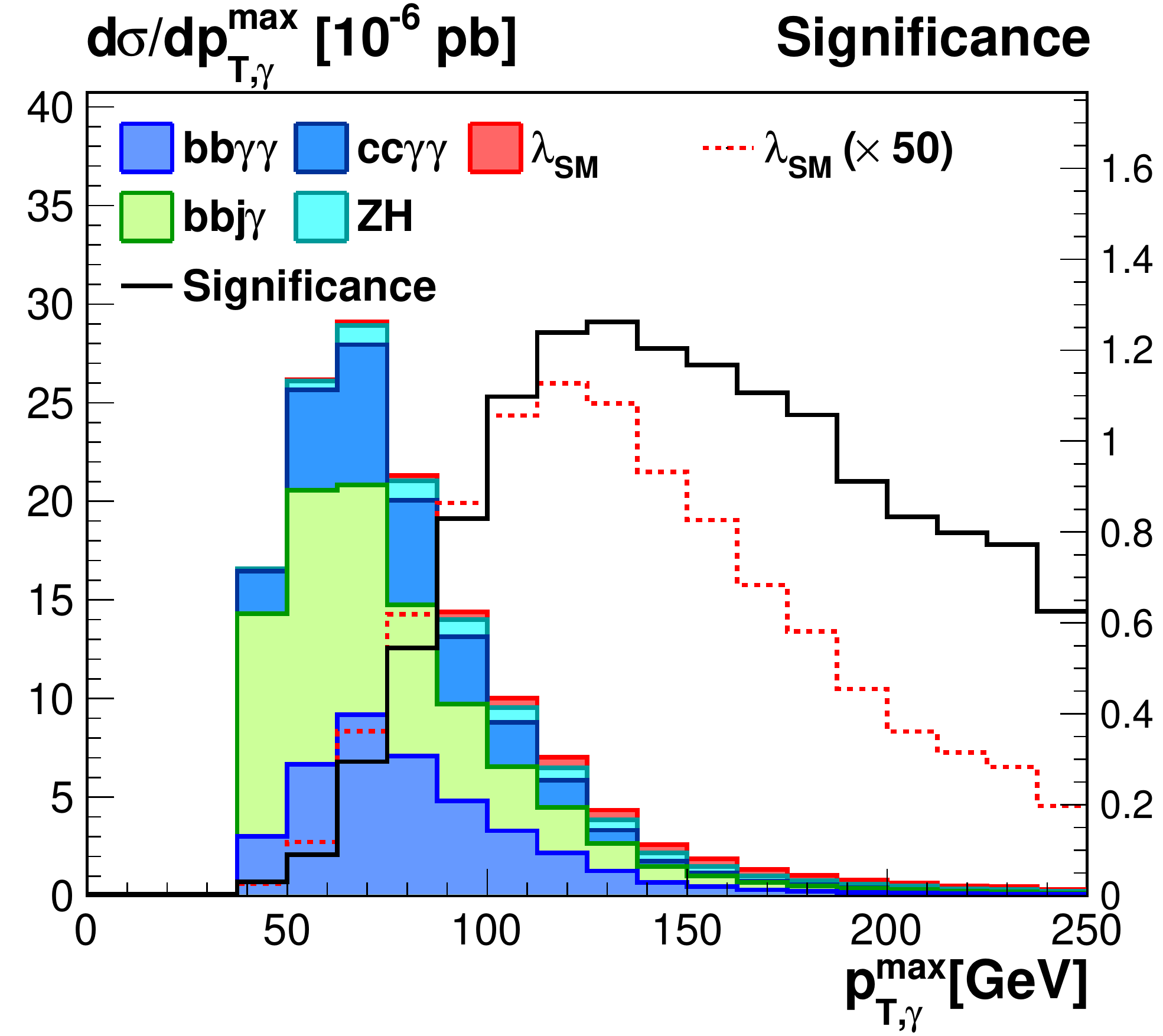}
  \hspace*{0.02\textwidth}
  \includegraphics[width=0.3\textwidth]{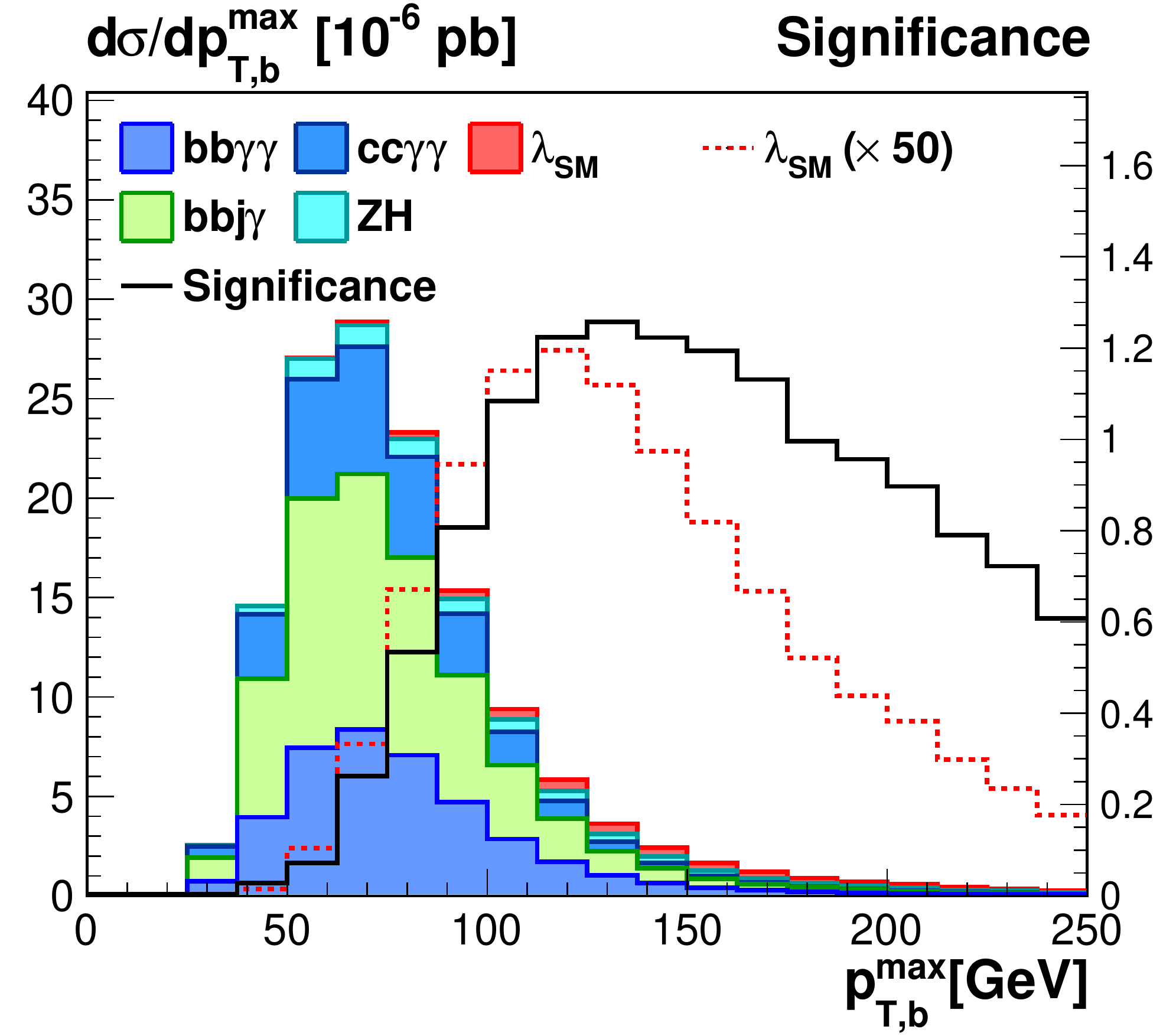}
  \caption{Kinematic distributions after trigger cuts for the SM signal
    (red) vs the $bb\gamma\gamma$, $bbj\gamma$, $cc\gamma\gamma$, and $ZH$
    backgrounds. The solid black line shows the differential
    distribution of the significance.}
\label{fig:background}
\end{figure}

In Fig.~\ref{fig:background} we show a set of kinematic distributions
for the Higgs decay products and for the reconstructed Higgs. At the
$2\to 2$ level we know that two distributions we want to study are the
invariant mass distribution $m_{HH}$ and the transverse momentum of
each of the two reconstructed Higgses. The dashed red line simply
scales the signal histogram so we can actually see its kinematic
distribution.  For both Higgs distributions the continuum backgrounds
are considerably softer than the Higgs pair signal. The same is true
for the transverse momenta of the Higgs decay products. From these
distributions it is also clear that an upper bound on $\Delta
R_{\gamma \gamma}$ will help us extract the Higgs pair
signal~\cite{uli3}.\bigskip

\begin{figure}[b!]
  \includegraphics[width=0.3\textwidth]{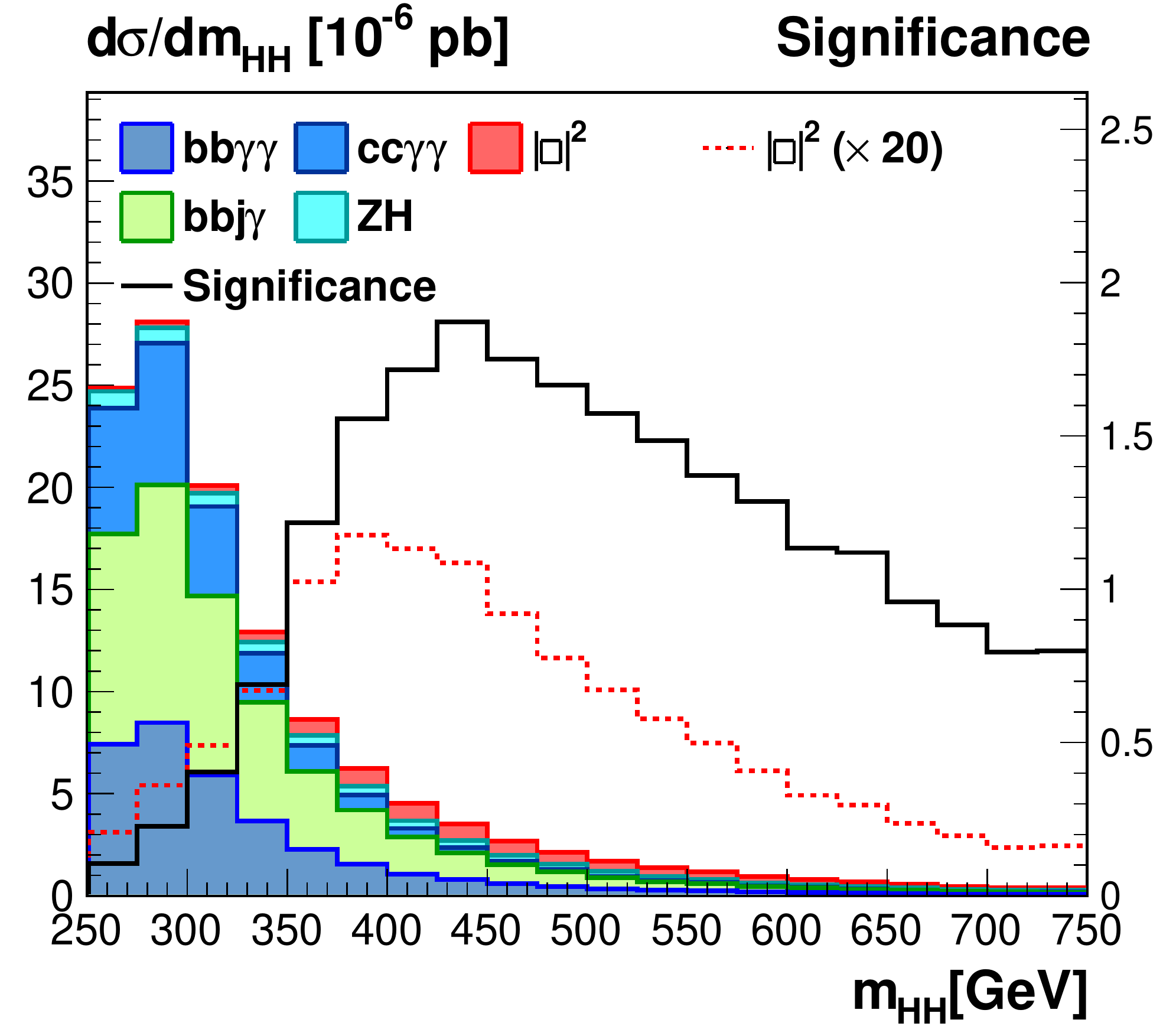}
  \hspace*{0.02\textwidth}
  \includegraphics[width=0.3\textwidth]{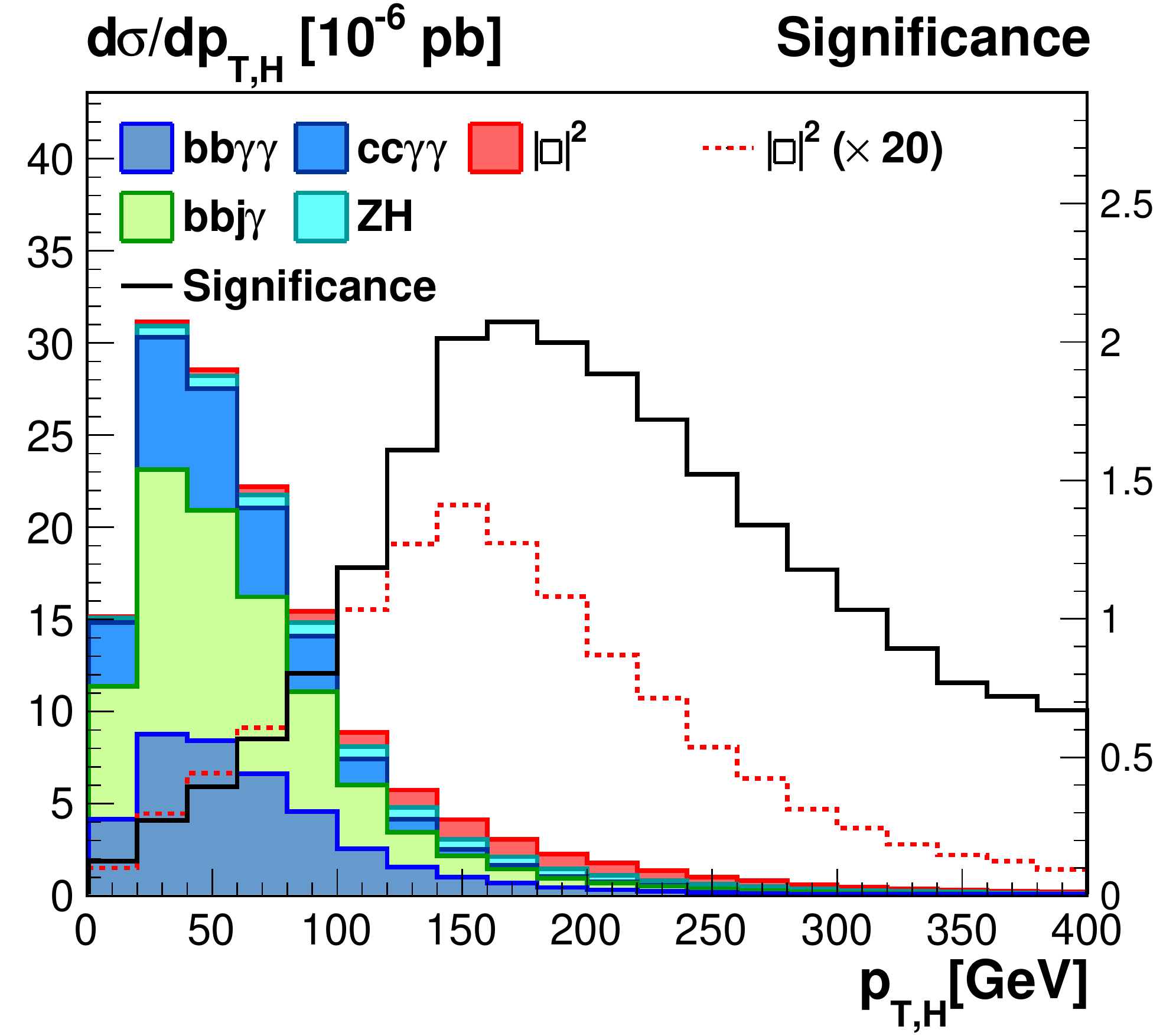}
  \hspace*{0.02\textwidth}
  \includegraphics[width=0.3\textwidth]{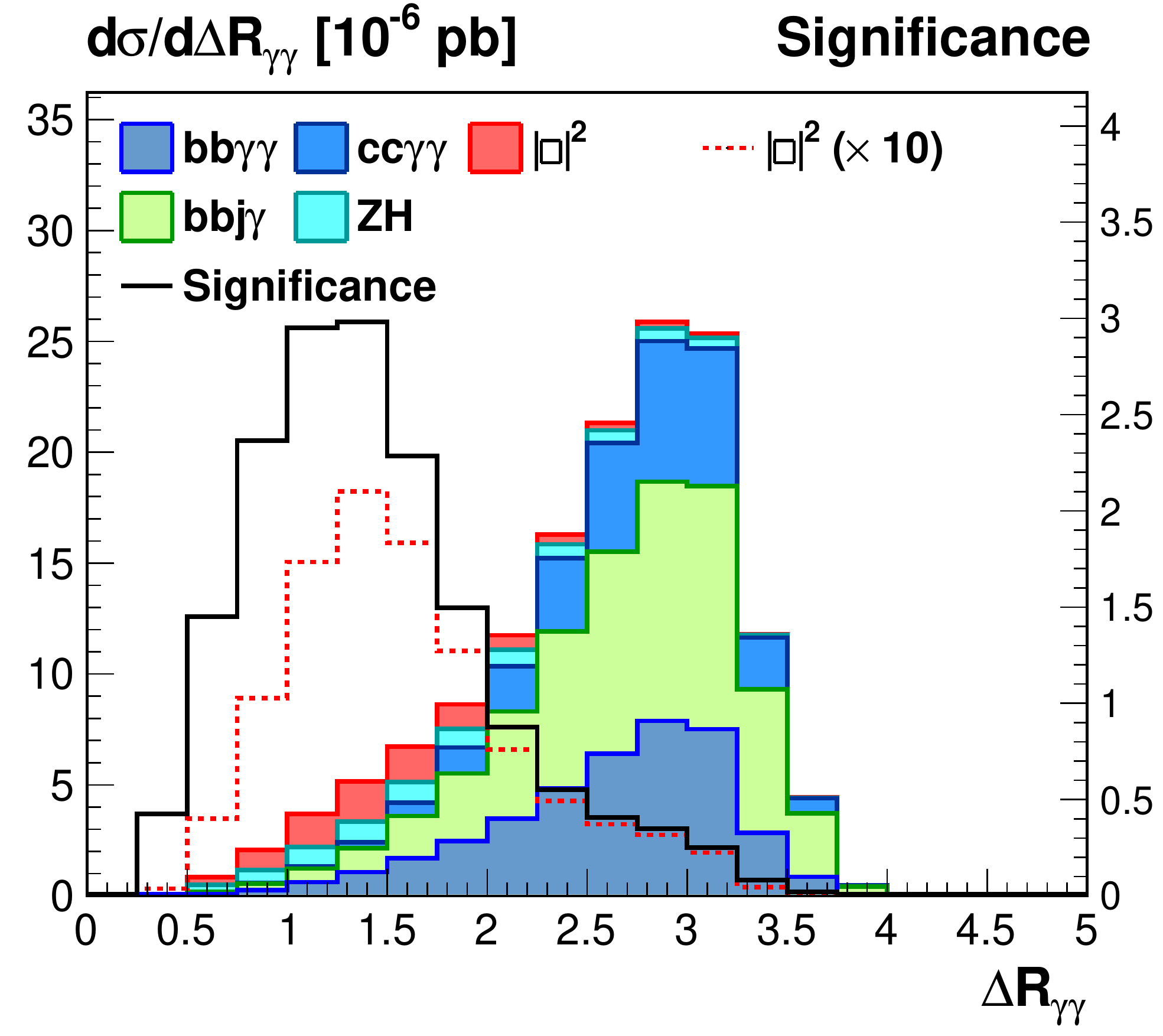}
  \includegraphics[width=0.3\textwidth]{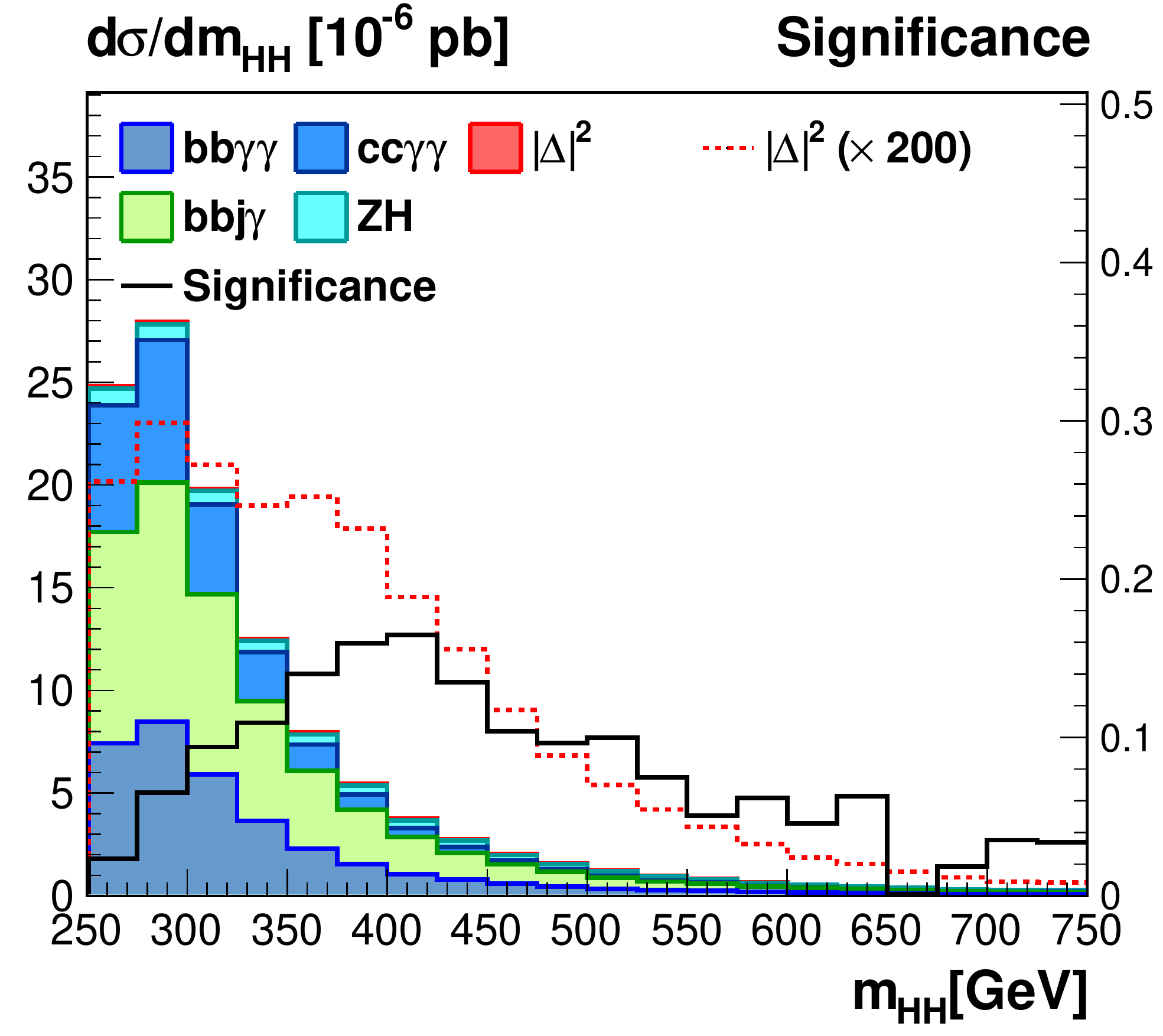}
  \hspace*{0.02\textwidth}
  \includegraphics[width=0.3\textwidth]{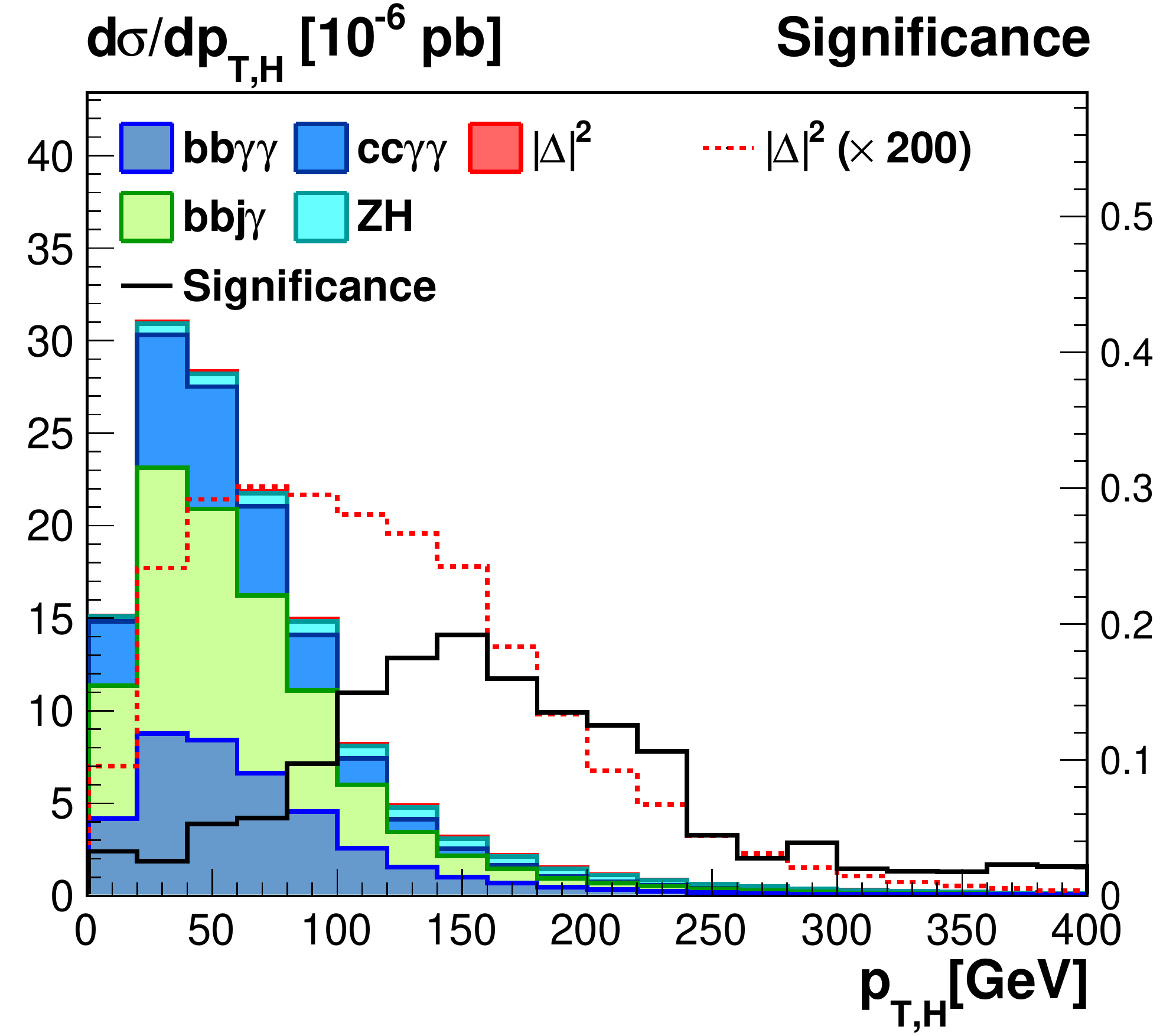}
  \hspace*{0.02\textwidth}
  \includegraphics[width=0.3\textwidth]{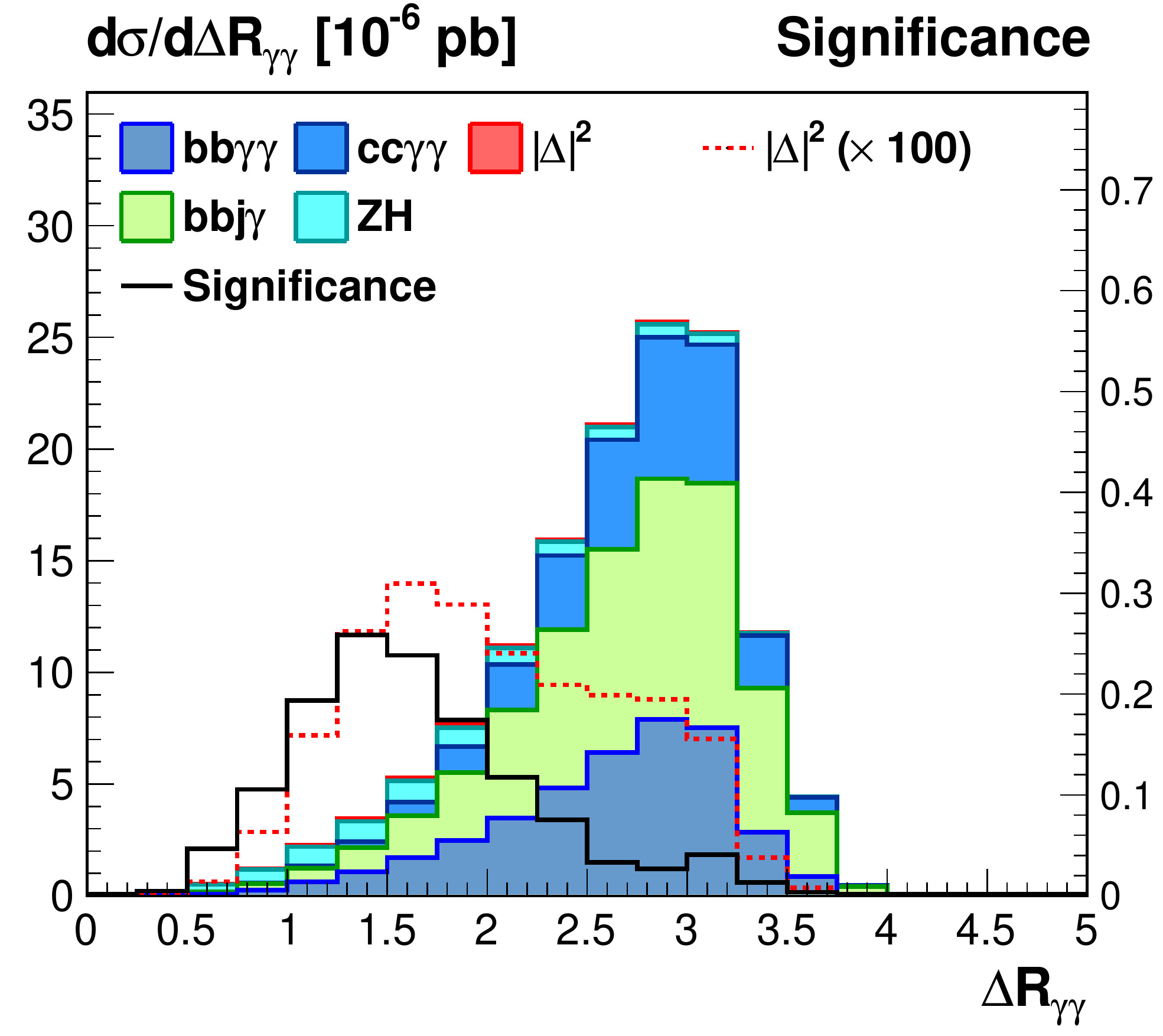}
  \caption{Kinematic distributions for the SM triangle diagram only
    (upper) and the SM box diagram only (lower).  We require trigger
    cuts for the signal (red) vs the $bb\gamma\gamma$, $bbj\gamma$,
    $cc\gamma\gamma$, and $ZH$ backgrounds. The solid black line shows
    the differential distribution of the significance.}
\label{fig:background_sep}
\end{figure}

For the maximum significance with which we can extract the Standard Model 
signal from the continuum background we find a Gaussian equivalent of
\begin{align}
4.02~\sigma \quad \text{for an integrated luminosity of} \; 3~\iab \; .
\end{align}
The black lines
in Fig.~\ref{fig:background} show how this significance is distributed
over phase space. First, we observe that the threshold region $m_{HH}
< 350$~GeV hardly contributes to the SM signal extraction. Instead, it
seems crucial to include reconstructed Higgses with $p_{T,H} >
150$~GeV, \ie well in the boosted regime~\cite{boosted,madmax2}. The
separation of the significance distribution from the signal
distributions occurs because of the rapidly falling background
distributions.  Due to the Higgs boost, widely separated photons with $\Delta R_{\gamma
  \gamma} > 1.8$ will not help with the signal extraction, while in
particular for the harder of the two photons or bottoms we can
completely ignore the soft part of the spectrum.

One key question for any multi-variate analysis is if the phase space
regions which dominate the signal vs background separation will be
safe with respect to systematic and theoretical uncertainties. From
the transverse momentum spectra in Fig.~\ref{fig:background} we see
that soft photons with $p_{T,\gamma} < 50$~GeV play hardly any role in
separating the Higgs pair signal from the continuum
background. Similarly, for the tagged bottom jets the relevant range
is $p_{T,b} = 100~....~250$~GeV. In this range we do not expect jet
radiation and the related combinatorics to have a large effect on our
results; the size of the usual perturbative rate corrections for the
signal and background process should be theoretically under control. 
Finally, the relevant di-photon phase space is $\Delta R_{\gamma
  \gamma} = 0.5~...~2$, clearly not a challenge for example to photon
separation.  A multi-variate analysis of Higgs pair production in the
Standard Model should be straightforward.\bigskip

From Eq.\eqref{eq:higgs_pair} we know that the two Feynman diagrams
contributing to Higgs pair production have distinct kinematic
features, and their combination should allow us to understand the
signal and significance distributions. In the upper panels of
Fig.~\ref{fig:background_sep} we illustrate some of the kinematic
distributions shown in Fig.~\ref{fig:background}, but for continuum
Higgs pair production only. As expected, the two cases are similar,
because the continuum diagram is responsible for almost the entire
signal rate in the Standard Model. In the second set of plots we show
the kinematic distributions of the triangle diagram only, \ie the term
of the cross section which carries the information on the Higgs
self-coupling. As a first piece of information, we see that the
significance with which we can extract the triangle diagrams in the
absence of an interference with the continuum is extremely
small. Kinematically, both in $m_{HH}$ and $p_{T,H}$ this diagram has
a considerably softer behavior. The two photons arising from a softer
Higgs decay can now be more widely separated. These are the phase
space regions where we can expect our signal vs background analysis to
gain sensitivity to the value of the Higgs self-coupling, typically 
through an enhanced cancellation between the two Feynman diagrams
shown in Fig.~\ref{fig:feynman}.

\section{Higgs self-coupling}
\label{sec:self}

\begin{figure}[t]
  \includegraphics[width=0.30\textwidth]{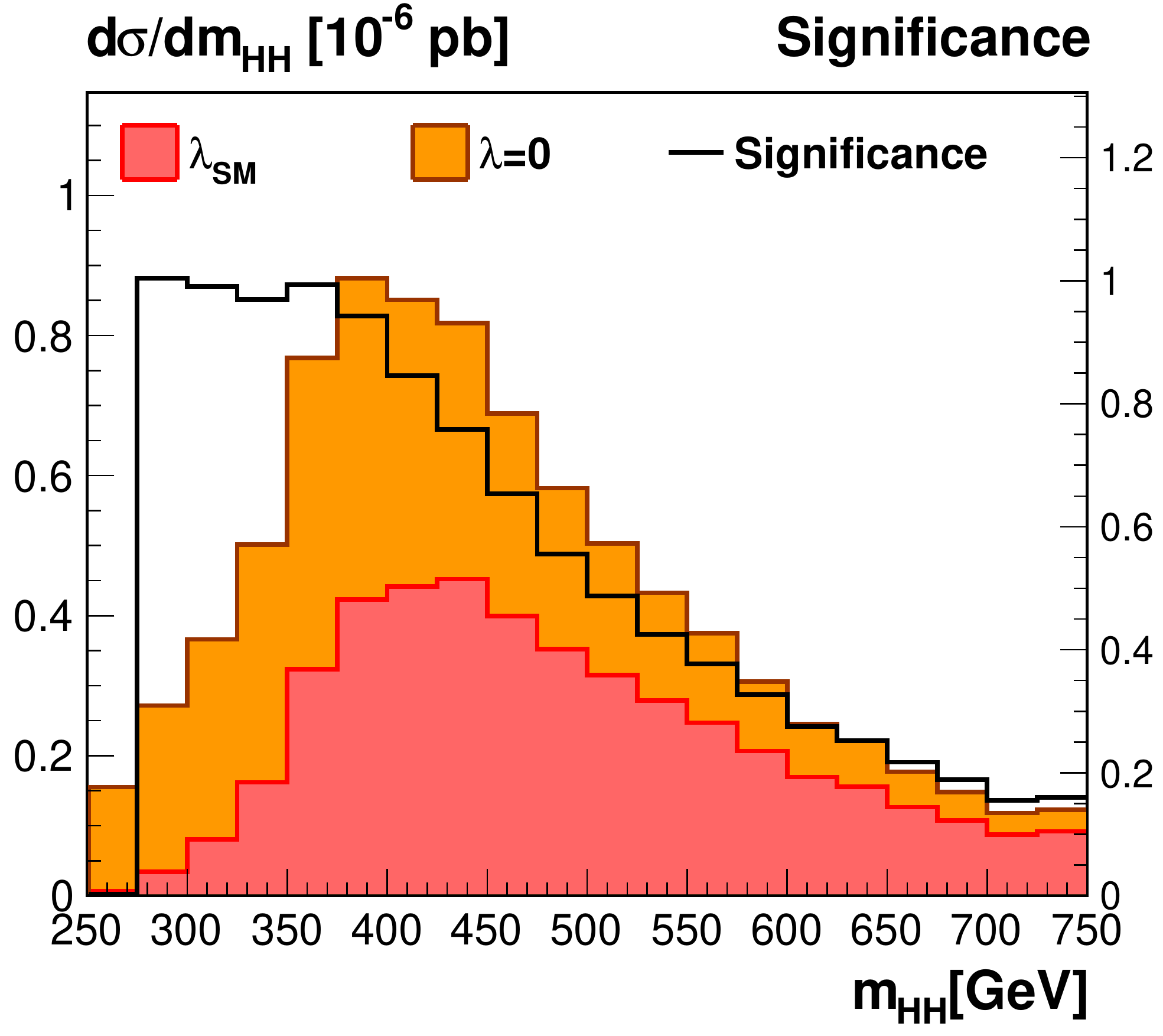}
  \hspace*{0.02\textwidth}
  \includegraphics[width=0.30\textwidth]{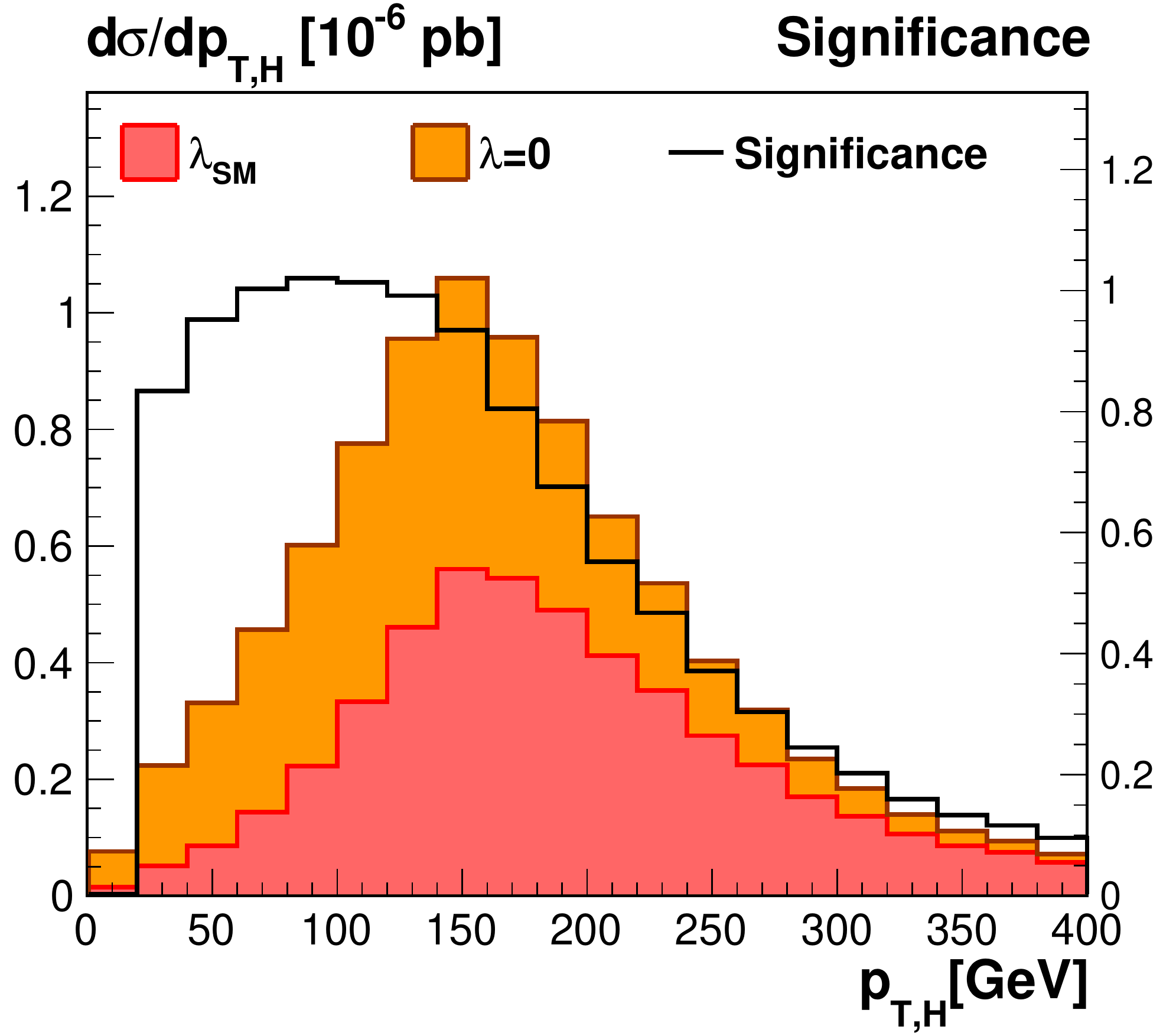}
  \hspace*{0.02\textwidth}
  \includegraphics[width=0.30\textwidth]{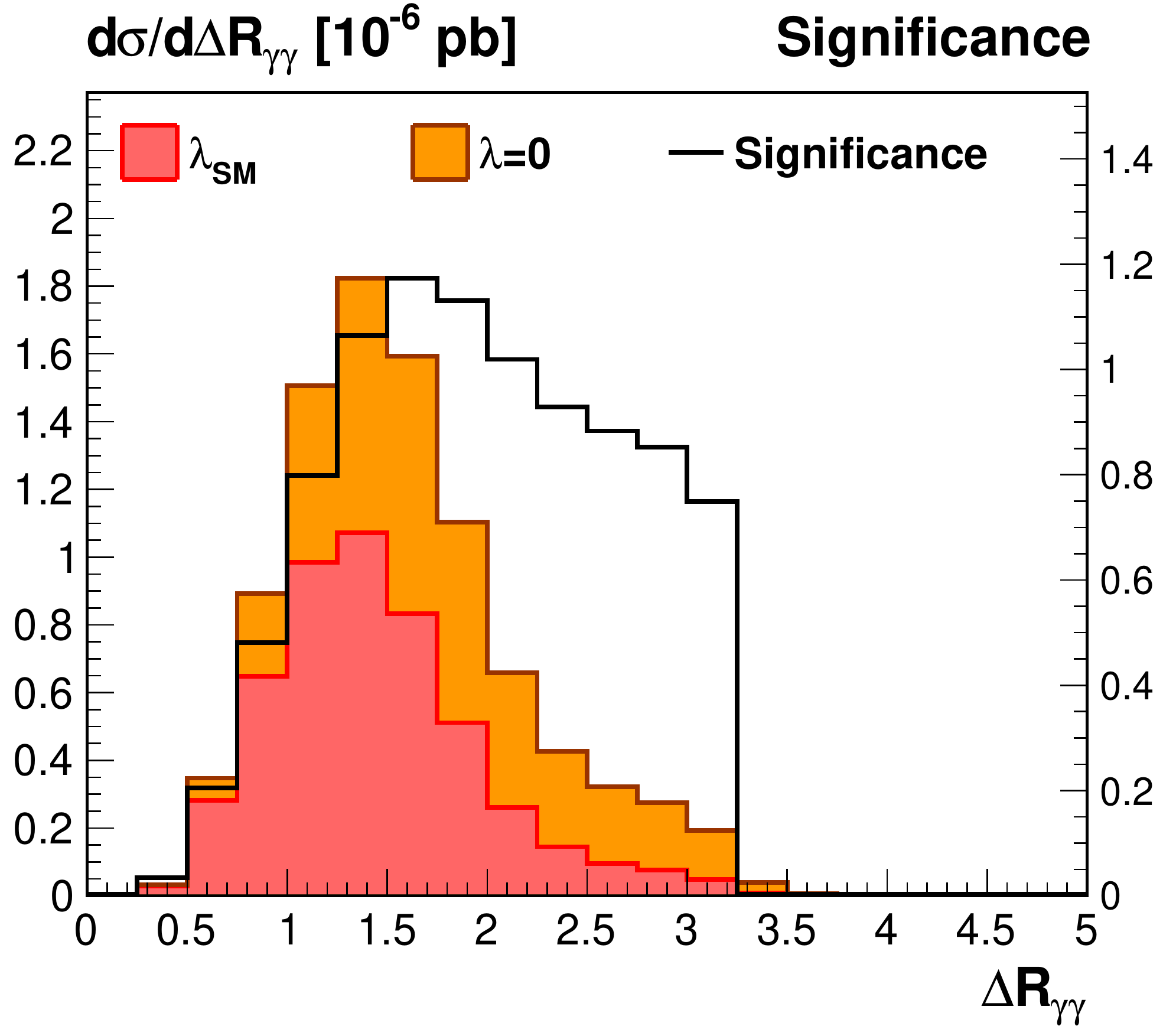} \\
  \includegraphics[width=0.30\textwidth]{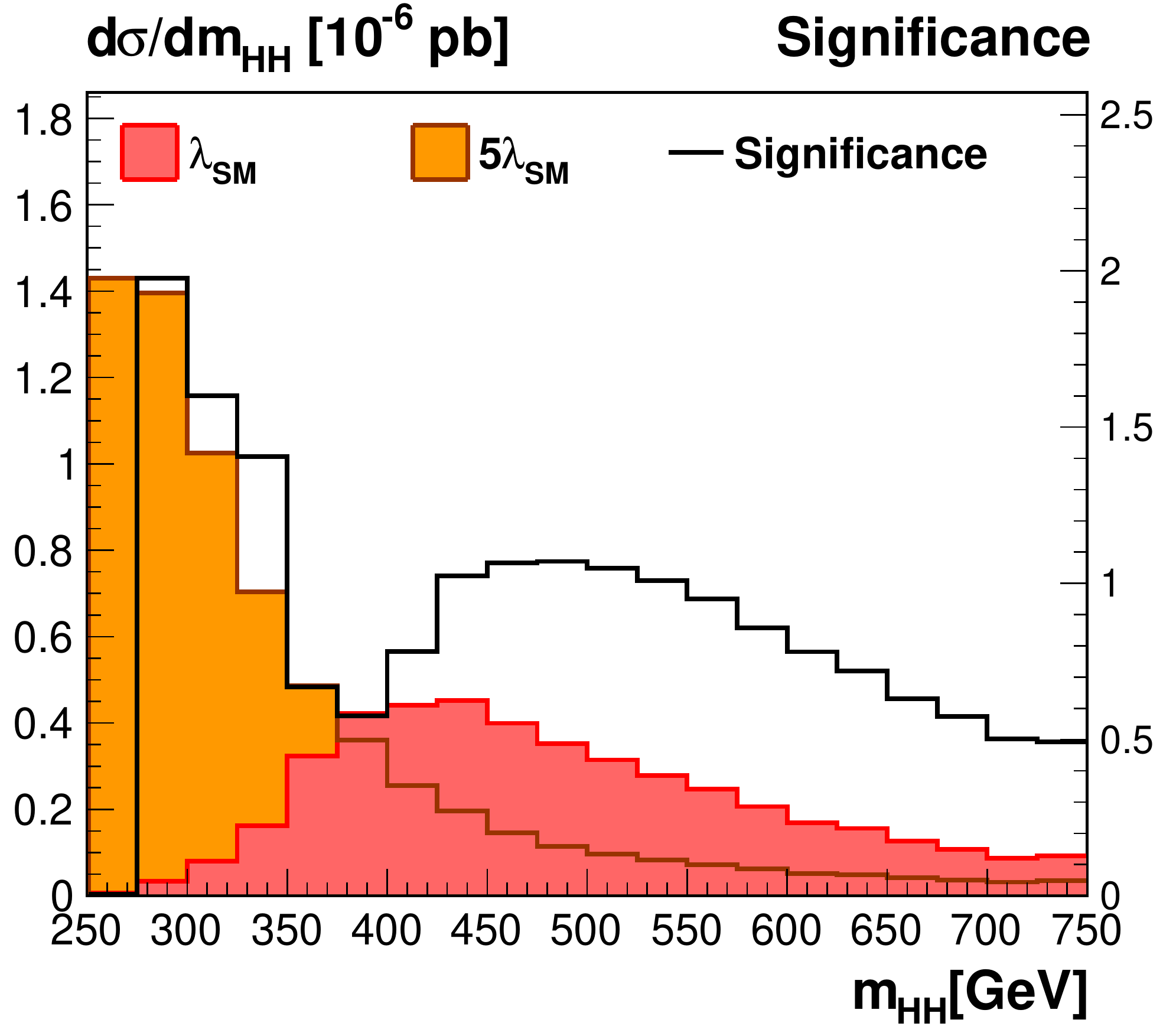} 
  \hspace*{0.02\textwidth}
  \includegraphics[width=0.30\textwidth]{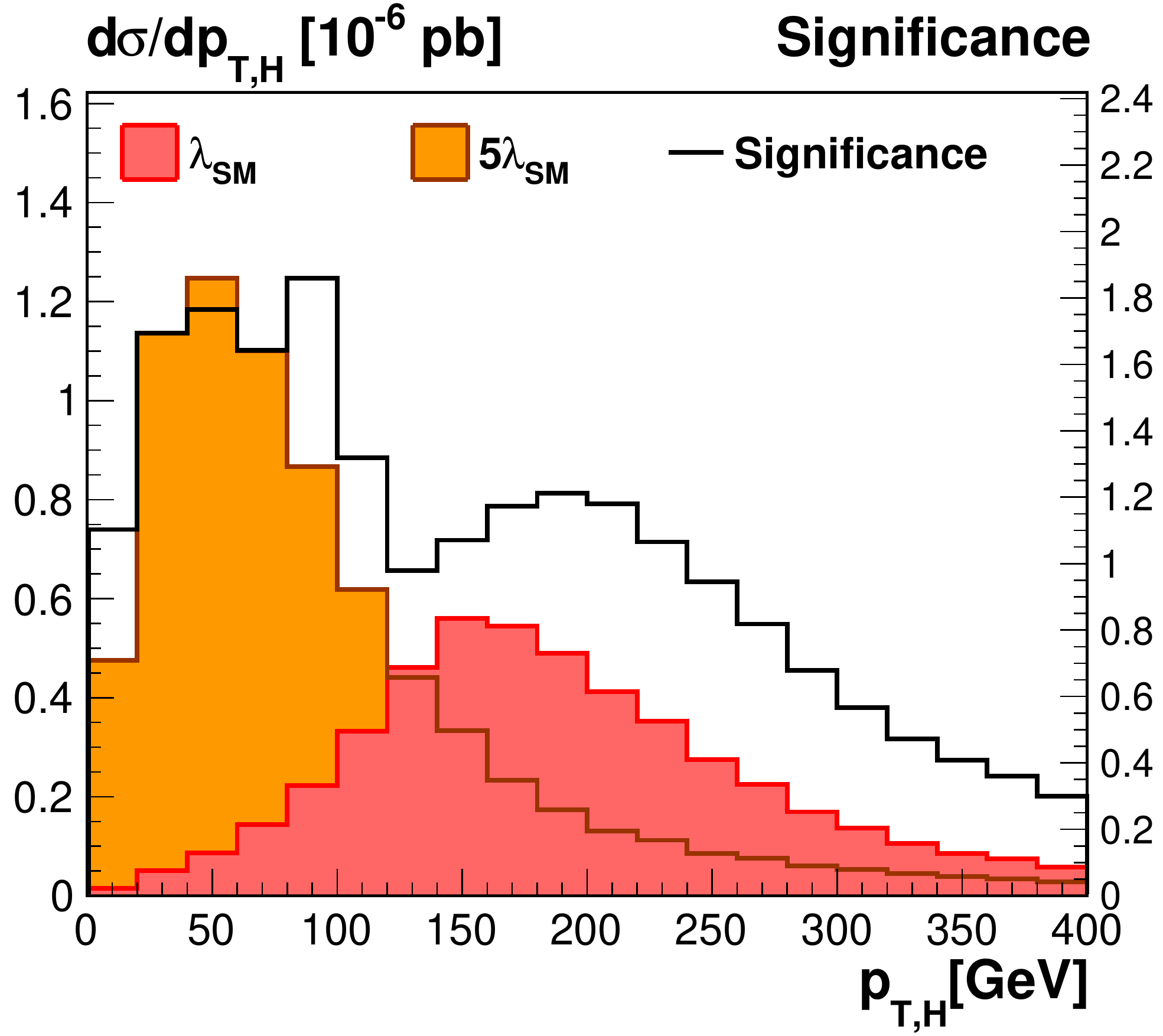} 
  \hspace*{0.02\textwidth}
  \includegraphics[width=0.30\textwidth]{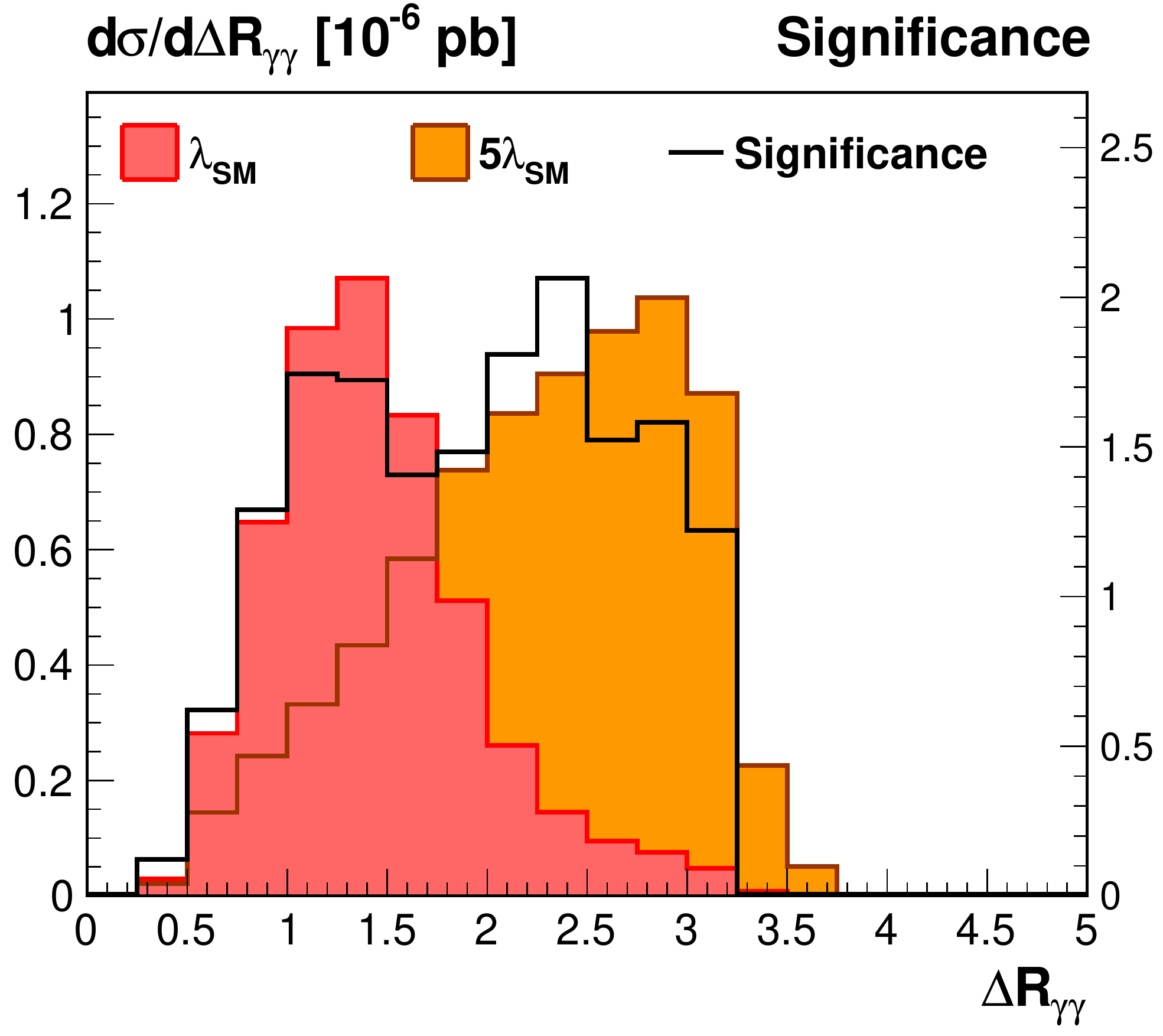} 
  \caption{Differential distributions assuming a modified Higgs
    self-coupling of $\lambda/\lambda_\text{SM}=0$ (upper panels) and $\lambda/\lambda_\text{SM} =5$ (lower panels).
    We compare only the two Higgs pair signals and neglect backgrounds
    for illustration.}
\label{fig:coup_s}
\end{figure}

\begin{figure}[b!]
  \includegraphics[width=0.40\textwidth]{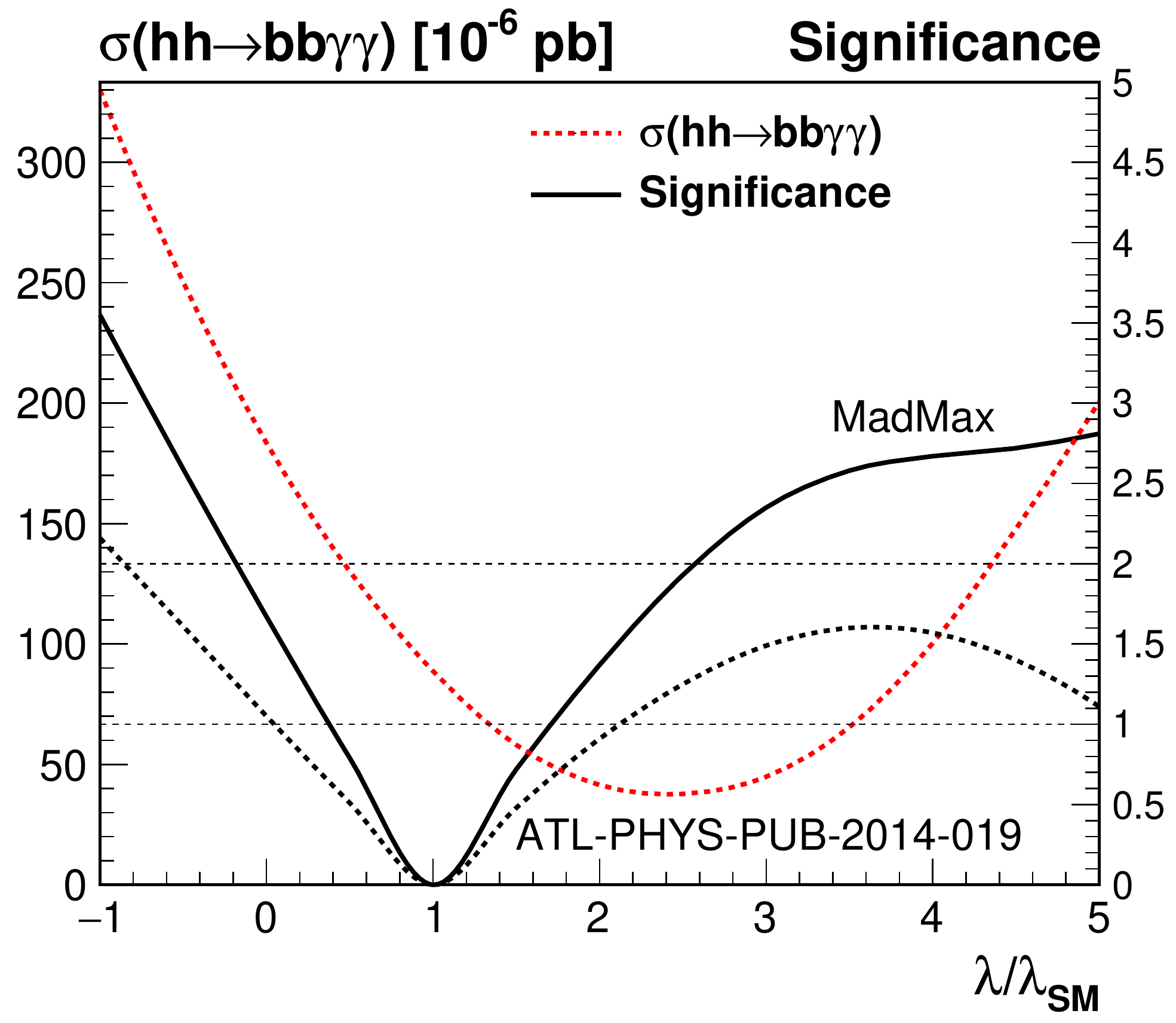}
  \caption{Signal cross section (red-dashed) and maximum significance
    (black-solid) for observing an anomalous Higgs self-coupling at
    the LHC with an integrated luminosity of $3~\iab$. We also show
    the significance from a cut-based rate measurement using the cuts
    suggested in Ref.~\cite{ATL-PHYS-PUB-2014-019} (black-dashed).}
\label{fig:total}
\end{figure}

In the second part of the paper we investigate the question:
\textsl{what are the prospects of measuring the Higgs self-coupling
  $\lambda$ in the presence of backgrounds?} In Fig.~\ref{fig:coup_s}
we show some kinematic distributions comparing two signal hypotheses,
assuming a variable self-coupling
\begin{align}
\frac{\lambda}{\lambda_\text{SM}} 
= \cfrac{\lambda}{\dfrac{3 m_H^2}{v}}
= 0, 2, 5 \; .
\end{align}
In the absence of a Higgs self-coupling we see that the Higgs pair
production rate is significantly enhanced. While it is not immediately
obvious from the two signal curves, the significance distribution over
$m_{HH}$ confirms that we can observe the effect of a zero
self-coupling mostly close to threshold and for $m_{HH} <
450$~GeV. Similarly, the absence of a self-coupling will modify the
transverse Higgs momentum around $p_{T,H} < 200$~GeV. In the photon
separation $R_{\gamma \gamma}$ the sensitive range requires widely
separated photons. We skip a detailed analysis of a slightly enhanced
self-coupling $\lambda = 2~\lambda_\text{SM}$, because it will be most
visible through a significant reduction of the production rate. The
dramatic case of $\lambda = 5~\lambda_\text{SM}$ is shown in the lower
panels of Fig.~\ref{fig:coup_s}. The kinematic distributions are now
modified close to threshold and for small transverse momenta of each
Higgs, as expected from an $s$-channel-mediated process $gg \to H^*
\to HH$.\bigskip

In the next step we add the QCD continuum
and $ZH$ backgrounds and test how well we can extract the Higgs self-coupling
from the kinematic information. In Fig.~\ref{fig:total} we show the
maximum significance of a multi-variate analysis as well as the
significance from a counting experiment based on the total Higgs pair cross section
after applying some basic cuts~\cite{ATL-PHYS-PUB-2014-019}. As
mentioned above, the Higgs pair production rate increases towards
vanishing self-coupling, has a minimum of roughly a quarter of the
Standard Model rate around $\lambda =2.3~\lambda_\text{SM}$, and
increases again for larger self-couplings. The arguably most
interesting case of zero self-coupling, related to a Higgs potential
without the usual minimum, should be most easily distinguishable from
the Standard Model~\cite{spirix,uli1}.  The LHC will at best constrain the Higgs self-coupling
range to be
\begin{align}
\frac{\lambda}{\lambda_\text{SM}} = 0.4~...~1.7 \quad \text{at 68\% CL and for} \; 3~\iab 
\end{align}
and rule out 
\begin{align}
\frac{\lambda}{\lambda_\text{SM}} = -0.2~...~2.6 \quad \text{at 95\% CL and for} \; 3~\iab \; .
\end{align}
Obviously, we can try to combine this result with
additional signatures to enhance the final reach of the
LHC~\cite{uli1,uli2,uli3,boosted}.\bigskip

\begin{figure}[t]
  \includegraphics[width=0.30\textwidth]{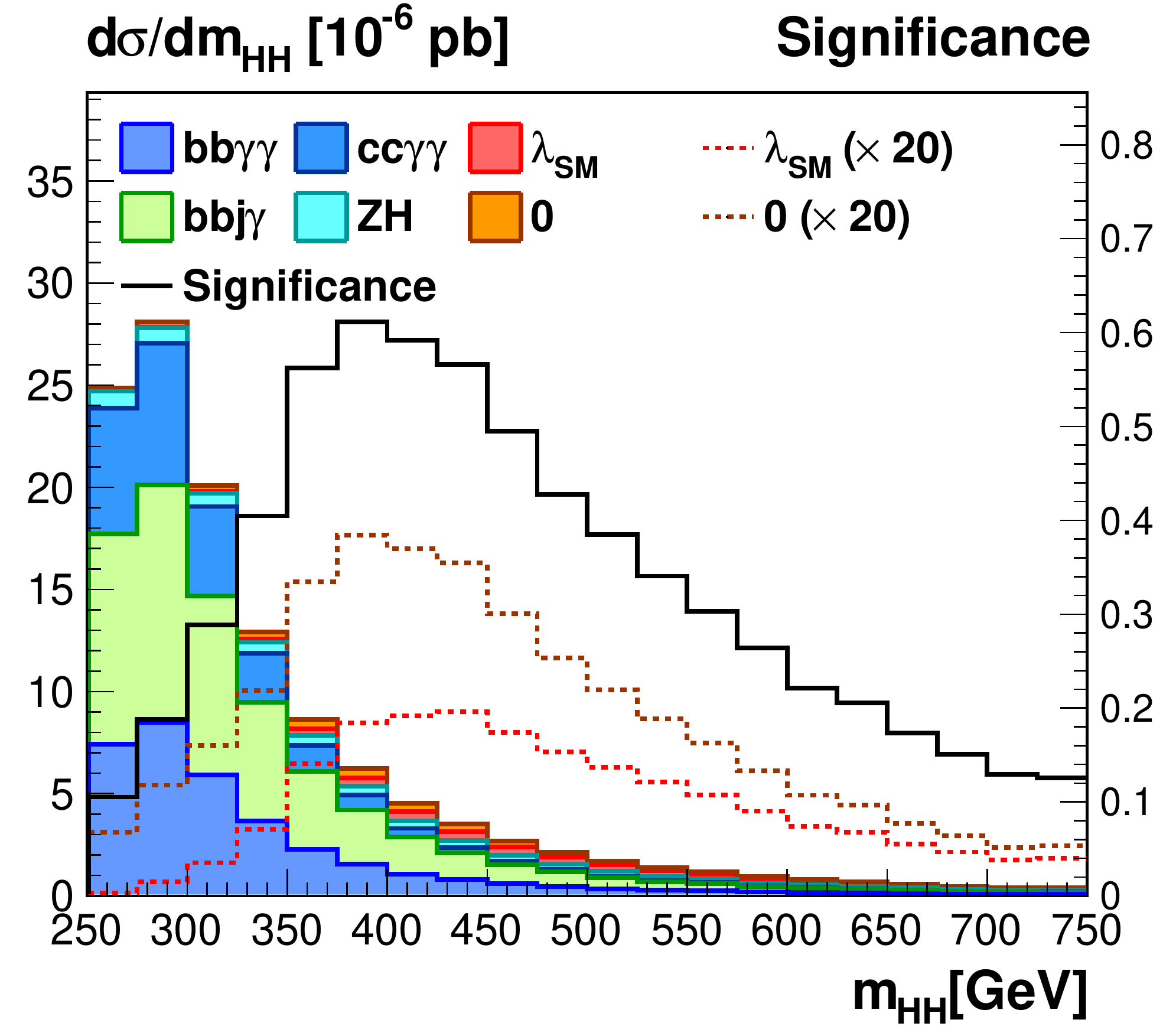}
  \hspace*{0.02\textwidth}
  \includegraphics[width=0.30\textwidth]{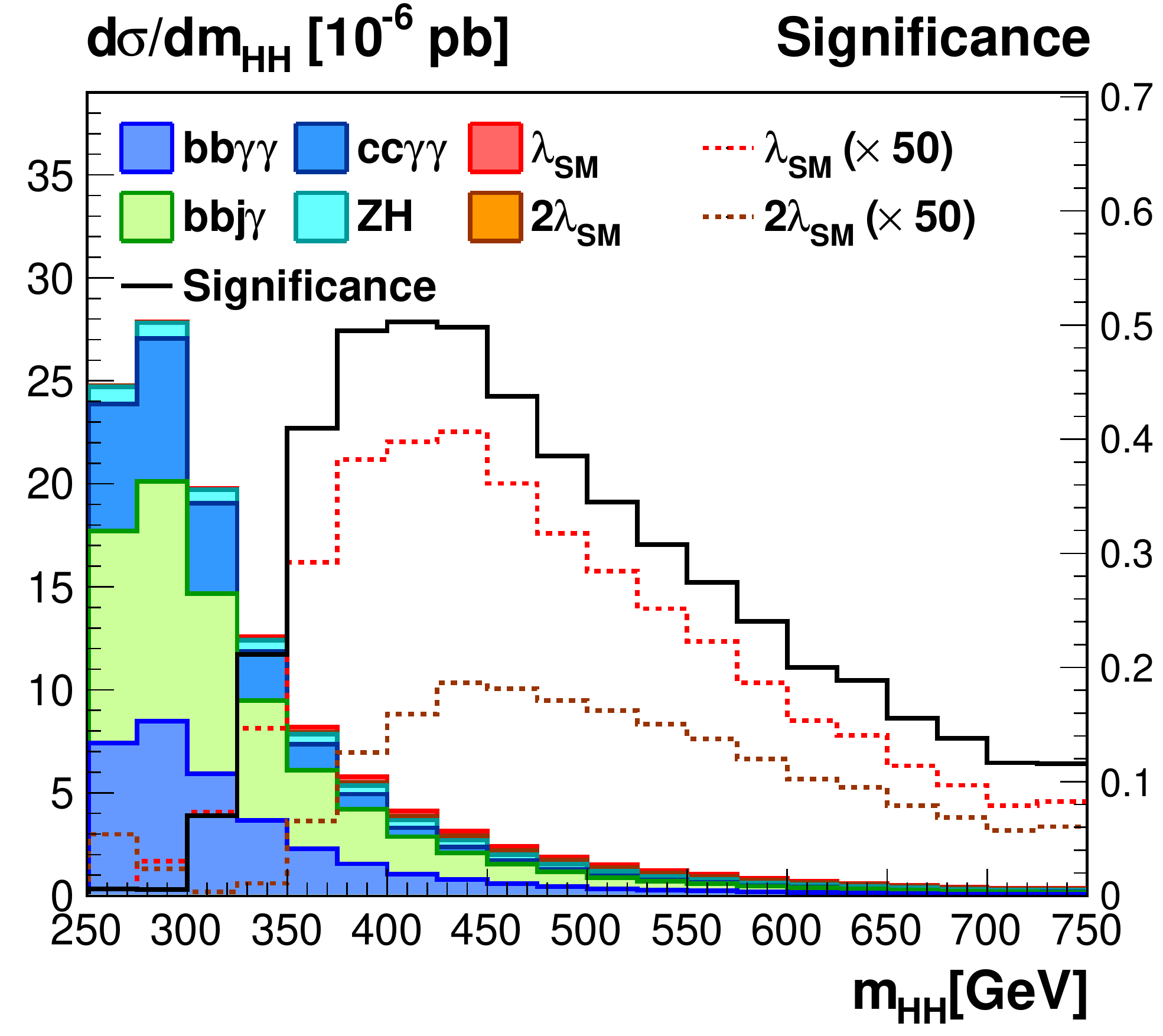}
  \hspace*{0.02\textwidth}
  \includegraphics[width=0.30\textwidth]{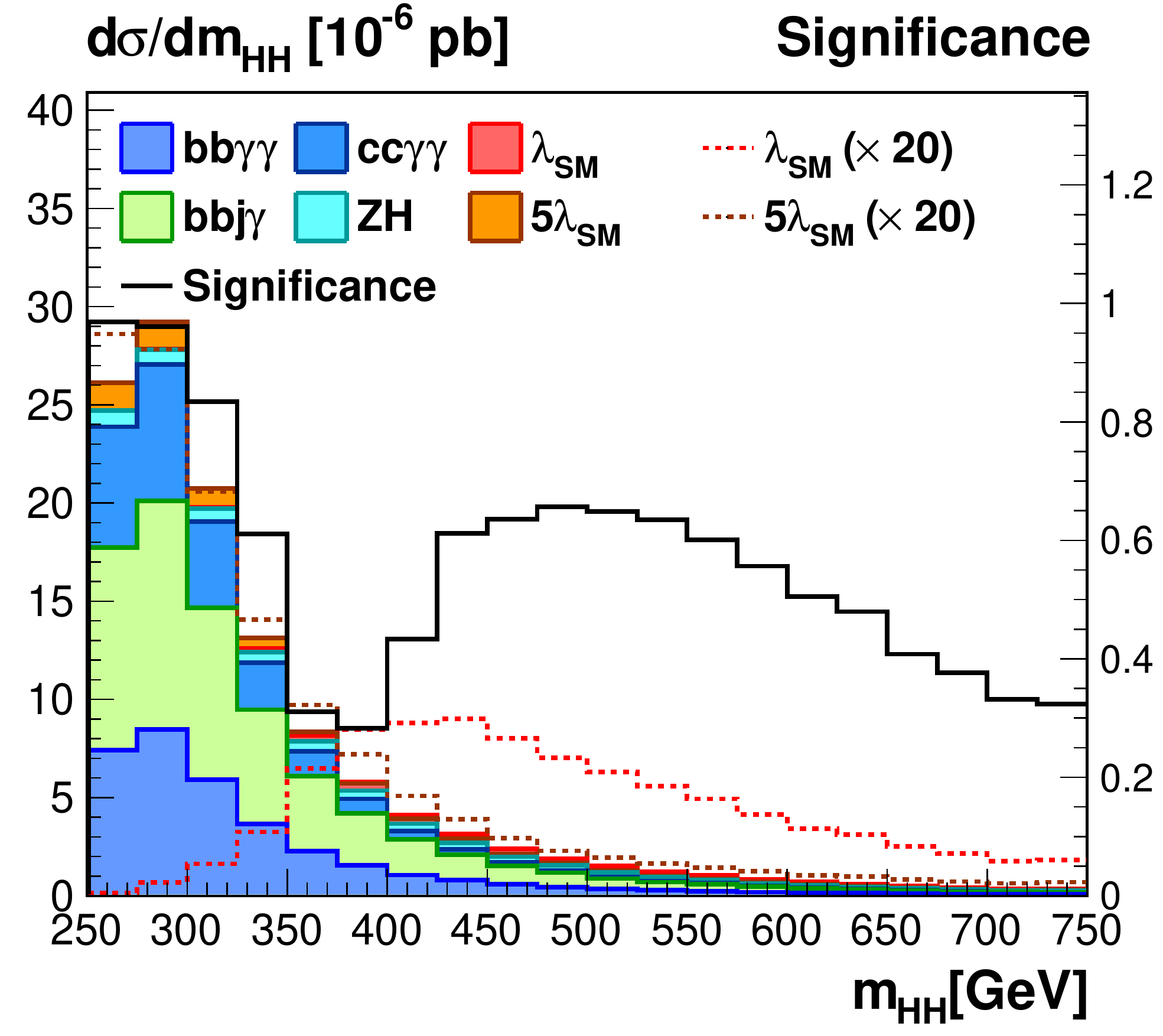} \\
  \includegraphics[width=0.30\textwidth]{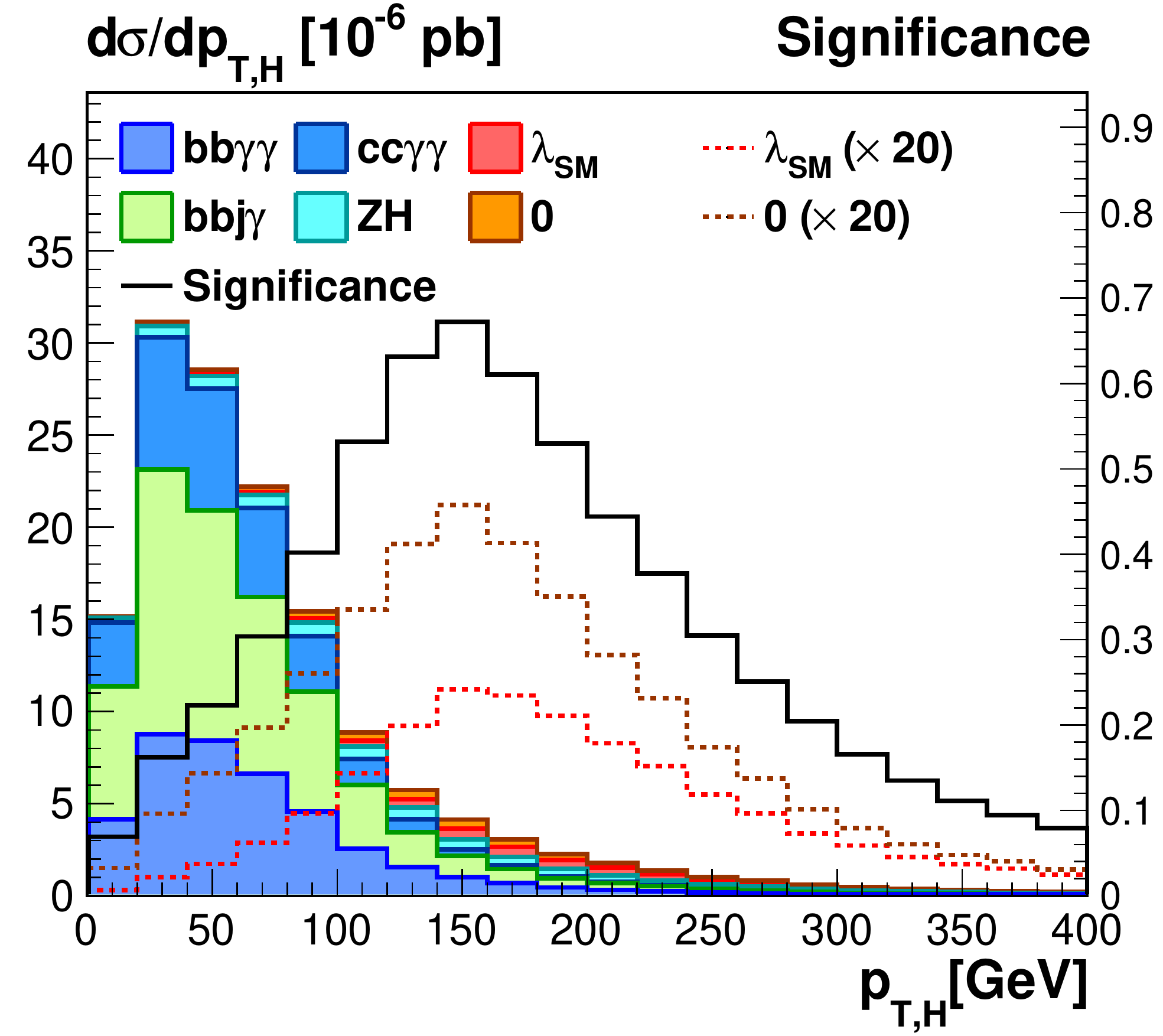}
  \hspace*{0.02\textwidth}
  \includegraphics[width=0.30\textwidth]{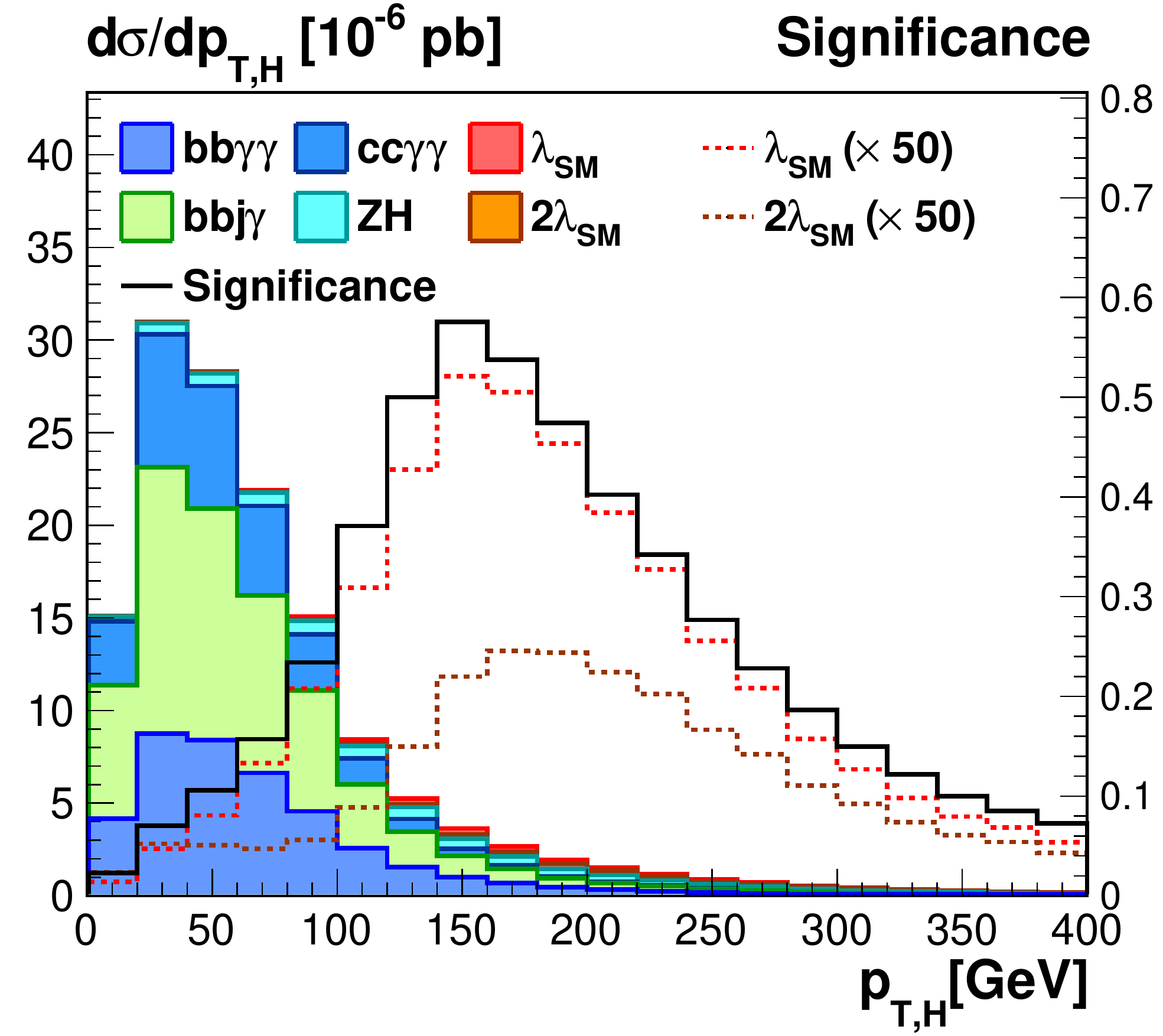}
  \hspace*{0.02\textwidth}
  \includegraphics[width=0.30\textwidth]{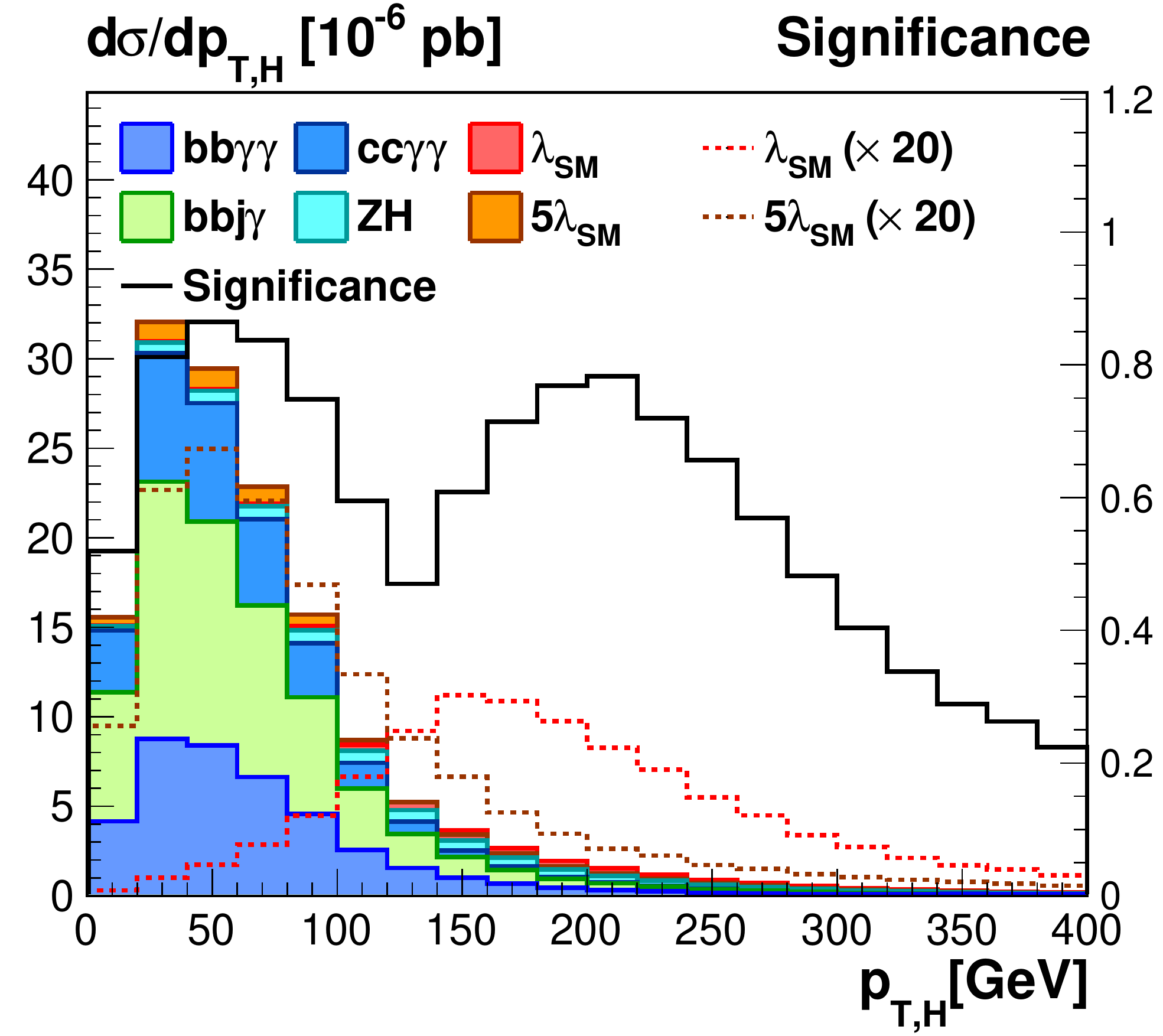} \\
  \includegraphics[width=0.30\textwidth]{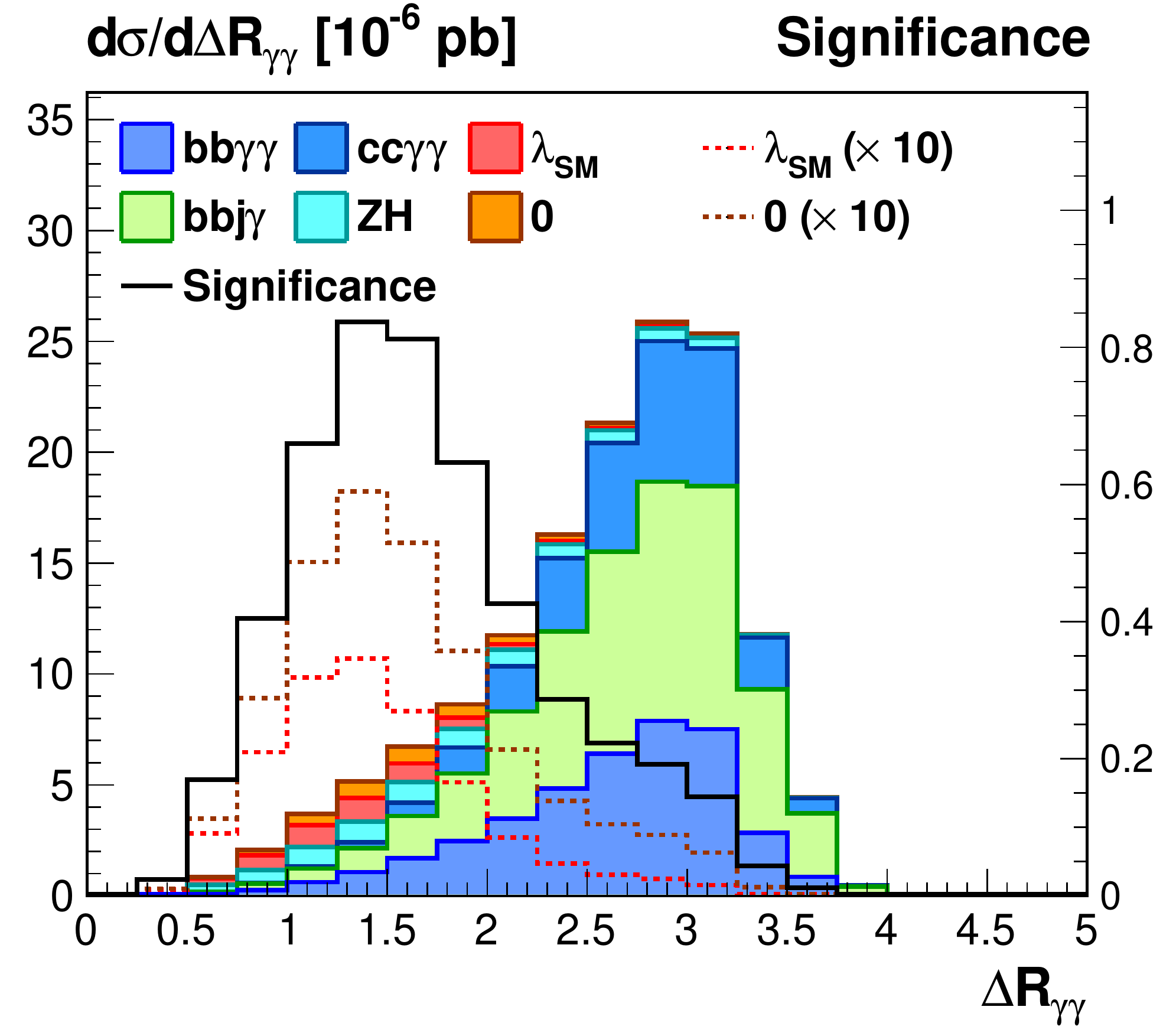}
  \hspace*{0.02\textwidth}
  \includegraphics[width=0.30\textwidth]{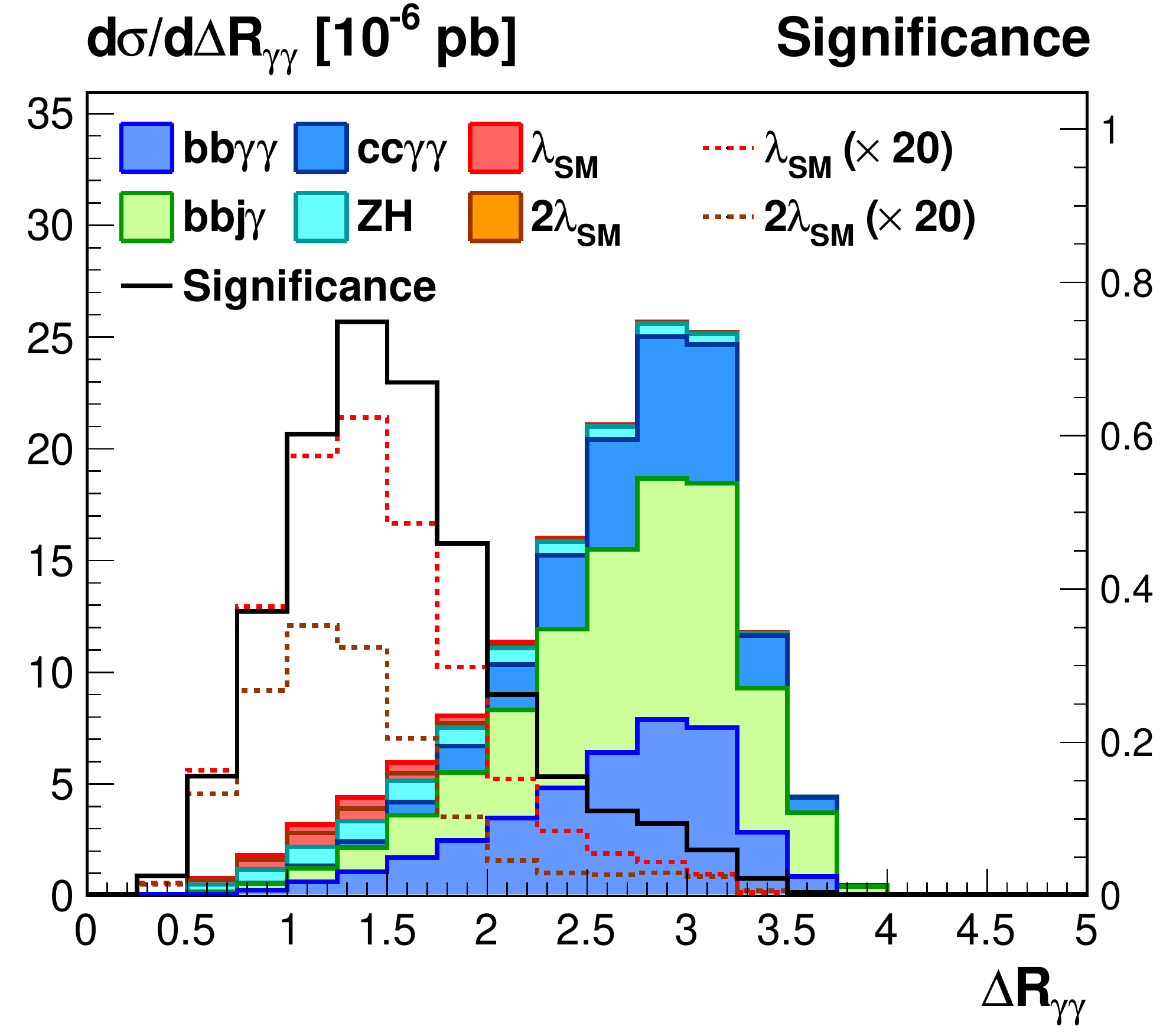}
  \hspace*{0.02\textwidth}
  \includegraphics[width=0.30\textwidth]{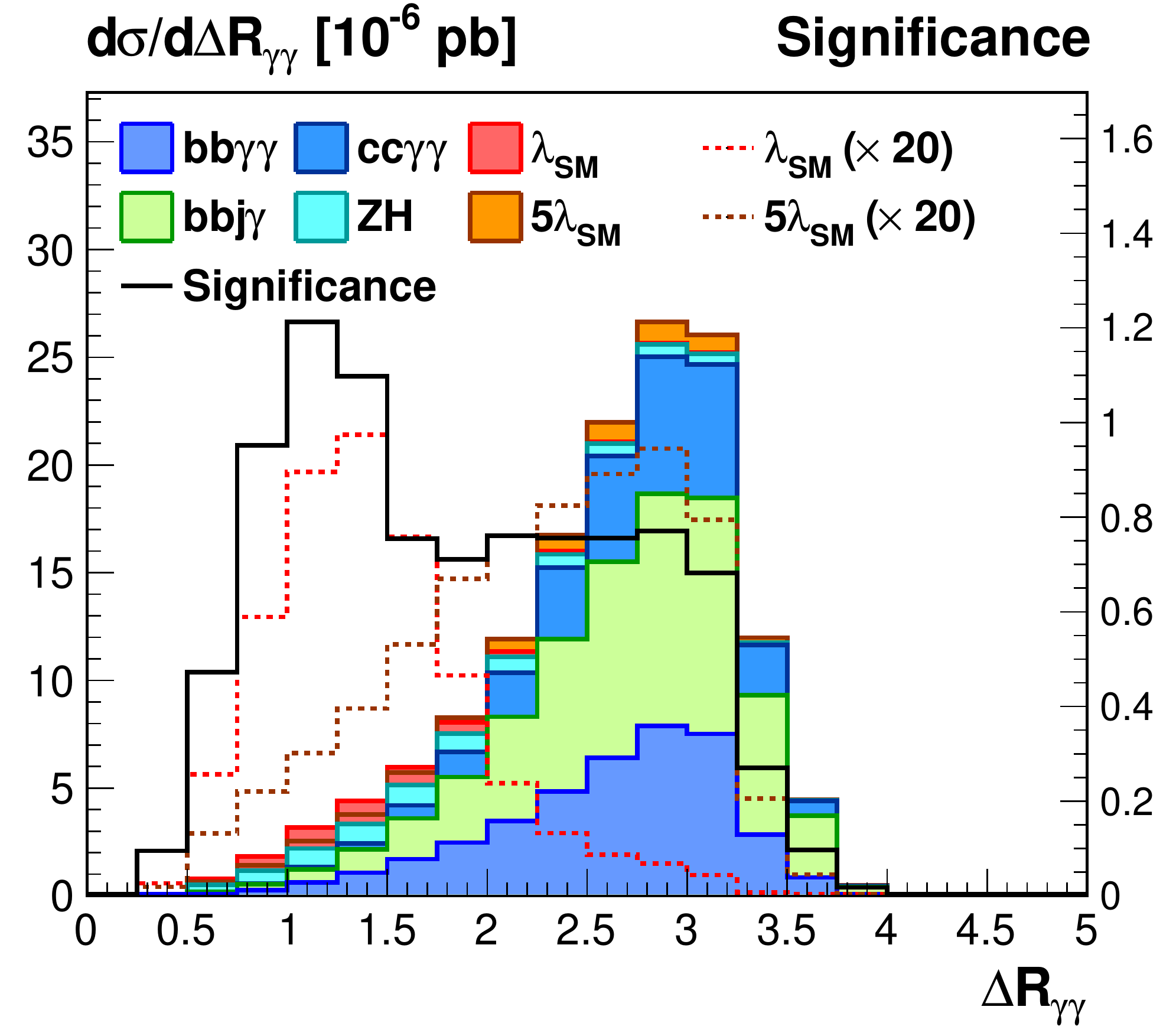} \\
  \includegraphics[width=0.30\textwidth]{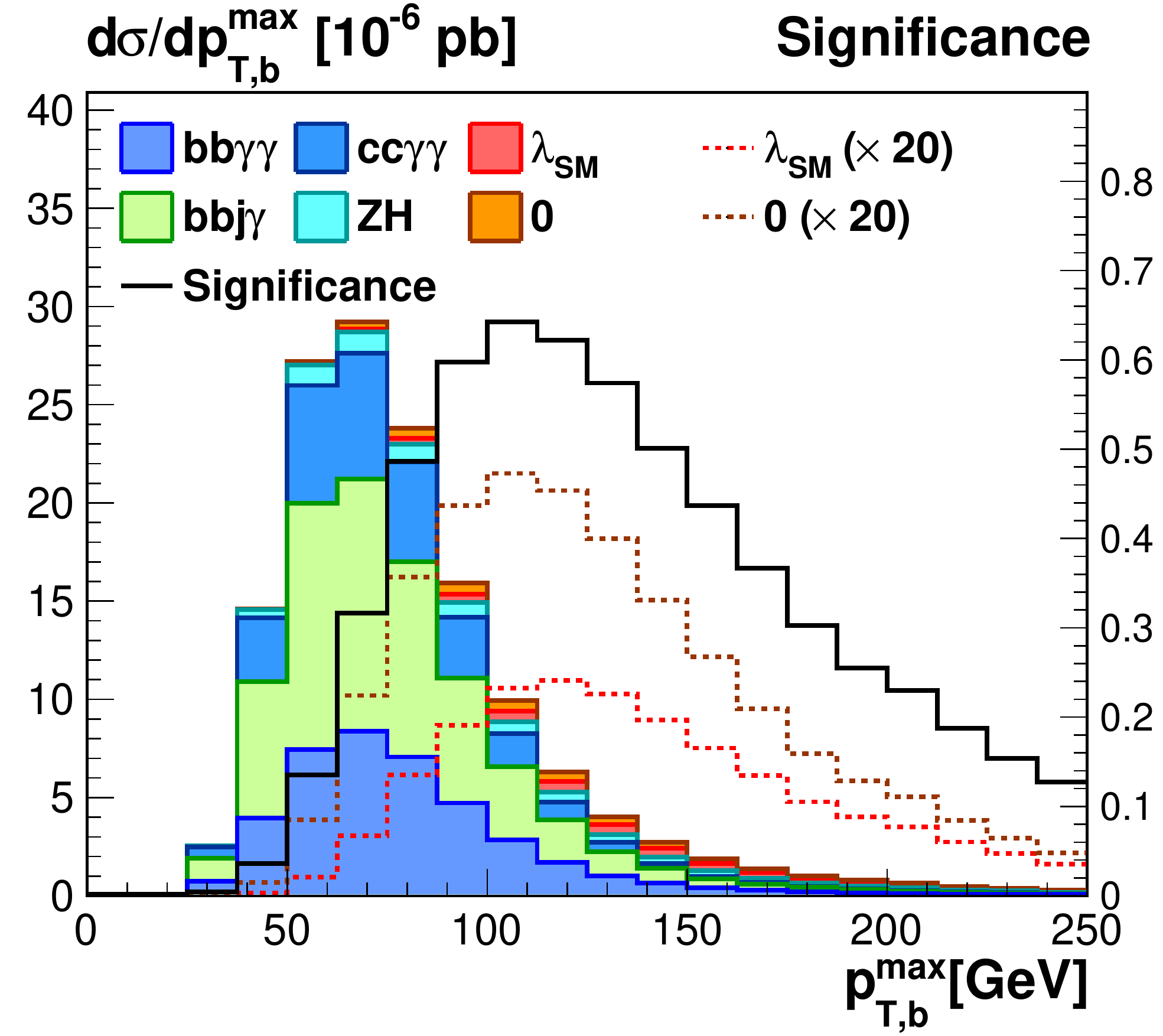}
  \hspace*{0.02\textwidth}
  \includegraphics[width=0.30\textwidth]{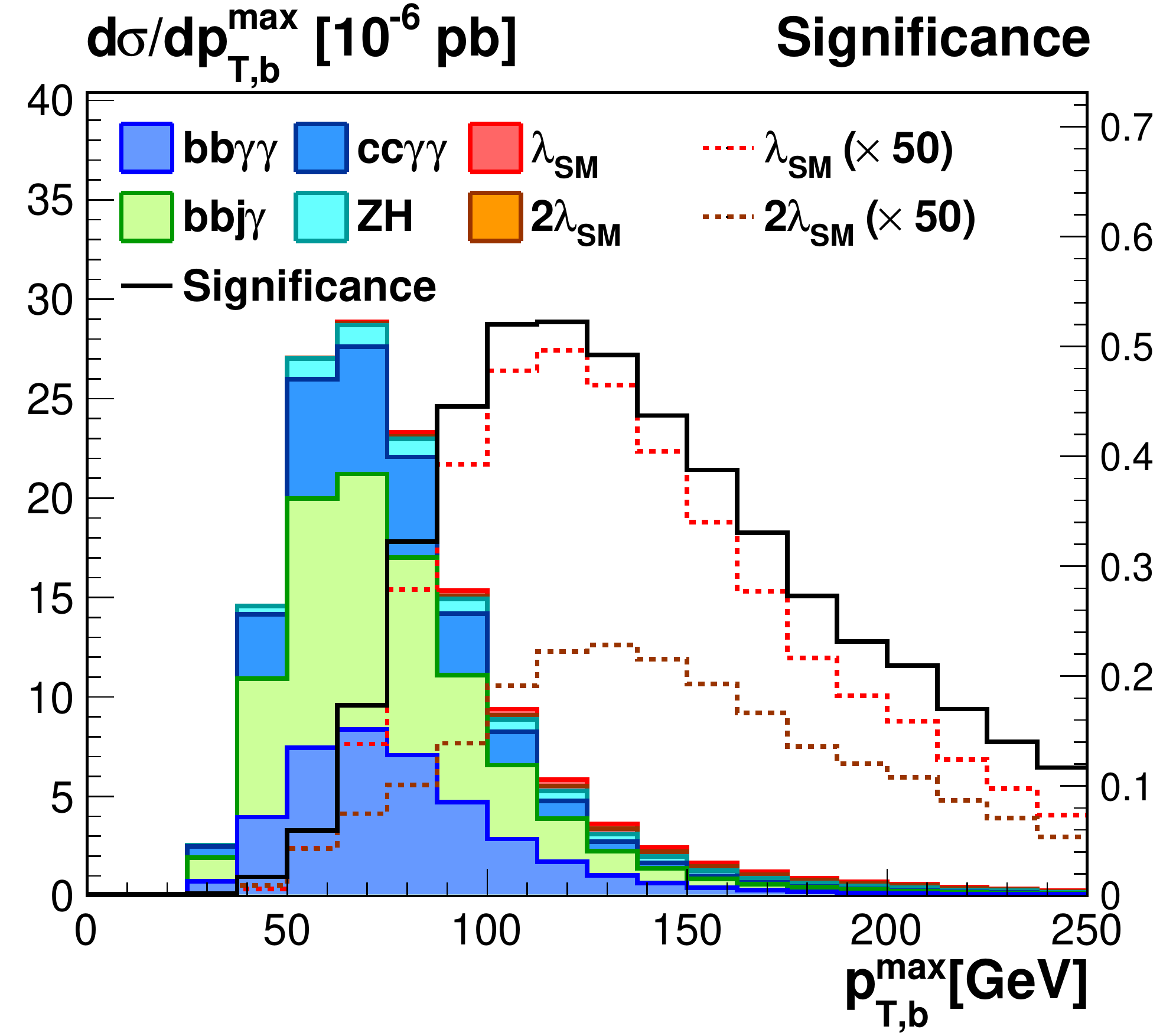}
  \hspace*{0.02\textwidth}
  \includegraphics[width=0.30\textwidth]{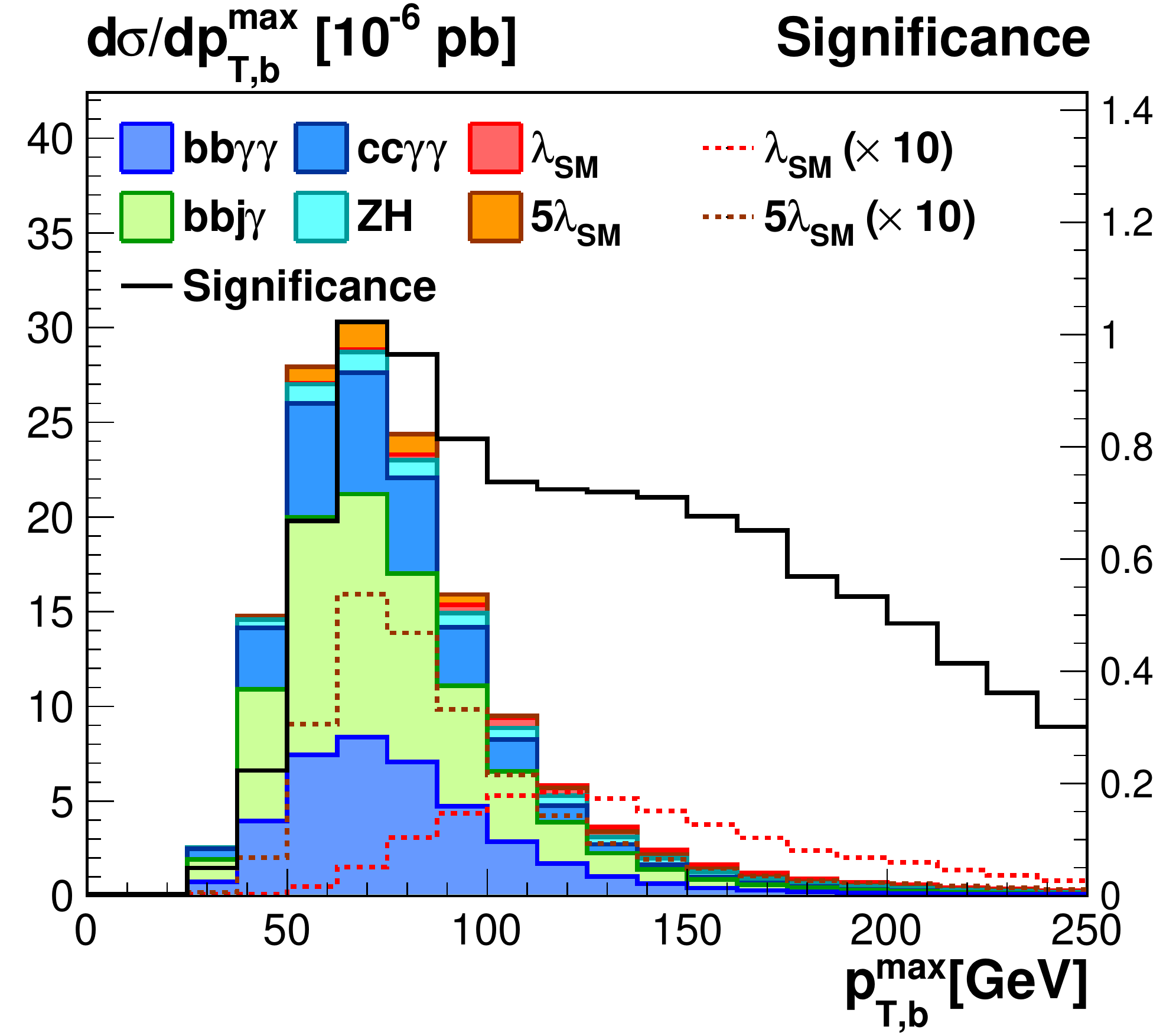}
  \caption{Differential distributions assuming a modified Higgs
    self-coupling of $\lambda/\lambda_\text{SM}=0,2,5$ (left to
    right). Unlike for the illustration in Fig.~\ref{fig:coup_s} we
    compare the two proper hypotheses including signal and
    backgrounds. }
\label{fig:coup_sb}
\end{figure}

From all we have discussed to here, the obvious question is how much
of the pure signal interference structure and its sensitivity to the
self-coupling survives in the presence of backgrounds, and to what
degree background uncertainties can mimic an anomalous self-coupling.
In Fig.~\ref{fig:coup_sb} we compare pairs of hypotheses for signal
plus background, \ie we test how well we can measure the Higgs
self-coupling in the presence of the backgrounds.  
The Standard Model signal is shown
in red and the alternative model with its shifted self-coupling
$\lambda/\lambda_\text{SM}=0,2,5$ is shown in orange. 
In our simulation we always apply mass
windows for the $b\bar{b}$ and $\gamma \gamma$ pairs, which already
biases our background simulations towards signal-like, hard
configurations. Nevertheless, in Fig.~\ref{fig:coup_sb} we see that
even then the QCD backgrounds still populate low-$m_{HH}$ and
low-$p_{T,H}$ regions. As already shown in Fig.~\ref{fig:coup_s}, these are
exactly the phase space regions where the sensitivity to an anomalous
self-coupling is the largest.\bigskip

For the measurement of $\lambda$ in the presence of backgrounds,
the sensitive regions of phase space are defined by a combination of
background rejection and sensitivity to $\lambda$. While for
background rejection described in Fig.~\ref{fig:background} the region
with high $m_{HH}>400$~GeV and $p_{T,H}>150$~GeV are most useful, the
self-coupling measurement requires lower-$m_{HH}$ and lower-$p_{T,H}$
bins, as seen in Fig.~\ref{fig:coup_s}. As a result, the significance
peak in the SM background rejection around $m_{HH} = 450$~GeV moves to
slightly below 400~GeV when we are interested in $\lambda$. Similarly,
the background-driven significance at $p_{T,H} > 180$~GeV and the
self-coupling-sensitive region $p_{T,H} < 150$~GeV together give a
distinct peak at 150~GeV for $\lambda / \lambda_\text{SM} = 0$ or
$\lambda / \lambda_\text{SM} = 2$. In contrast, for a very large
self-coupling $\lambda / \lambda_\text{SM} = 5$ the significance
receives contributions from two distinct regions of phase space,
$p_{T,H} \approx 50$~GeV and $p_{T,H} \approx 220$~GeV. The geometric
separation of the two photons and the transverse momentum of the
harder $b$-jet follow the same pattern. This means that the most
significant phase space regions for a measurement of the self-coupling
are driven by the background rejection, shifted by the well-known
regions of phase space carrying sensitivity to the
self-coupling. Large deviations from the phase space regions for
background rejection only occur when we test very large
self-couplings.

\clearpage
\section{Outlook}
\label{sec:outlook}

Multi-variate analyses often challenge our understanding of what
limiting factors of important measurements are. To gauge the
sensitivity for example of a Higgs self-coupling measurement to
different sources of uncertainties we need to understand where the
relevant phase space regions for a given measurement
are~\cite{madmax1,madmax2}.

In this paper we have studied the phase space regions which contribute
to the extraction of the Higgs pair production signal from the
continuum backgrounds, as well as those regions allowing for a
measurement of the Higgs self-coupling. We focus on the $HH \to
b\bar{b} \gamma \gamma$ signature~\cite{uli3}, but expect our results
to also hold for other channels with large continuum backgrounds. The
two relevant phase space regions for the signal extraction and the
coupling measurement are separate and in particular for the signal
extraction well understood in terms of systematic and theoretical
errors.

The most sensitive phase space region for extracting the self-coupling
is close to threshold, where we expect the QCD background to overwhelm
the Higgs pair signal. The main question will be how well we
understand those backgrounds and how much of this region can still be
used for the self-coupling measurement. Assuming SM-like
self-coupling, the bulk of the coupling-sensitive region in the
presence of QCD backgrounds is only slightly softer than the relevant
phase space for the extraction of the Standard Model signal. For large
self-couplings this region shifts significantly, forcing us to
consider a proper hypothesis test for a variable Higgs self-coupling.\bigskip

\begin{center} \textbf{Acknowledgments} \end{center}

First, we still remember to Uli Baur for coming up with this
measurement and for developing the first and original analyses.
Second, we are grateful to Kyle Cranmer for his continuous support and
to Bill Murray for his valuable comments concerning v1. Moreover, we
would like to thank Martin Jankowiak for his contributions during an
early phase of the project. We thank David Bourne and Anthony Yeates
for helpful discussion. TP and PS would like to thank the MITP for the
hospitality and support while this paper was finally written.  The
work of F.K.~was supported by US Department of Energy under
Grant~DE-FG02-04ER-41298.  F.K.~ also acknowledges support from the
Fermilab Graduate Student Research Program in Theoretical Physics
operated by Fermi Research Alliance, LLC under Contract
No. DE-AC02-07CH11359 with the United States Department of
Energy. P.S. work was supported in part by the European Union as part
of the FP7 Marie Curie Initial Training Network MCnetITN
(PITN-GA-2012-315877).

\newpage

\appendix

\section{Future 100~TeV Collider}
\label{app:100}

\begin{figure}[t]
  \includegraphics[width=0.40\textwidth]{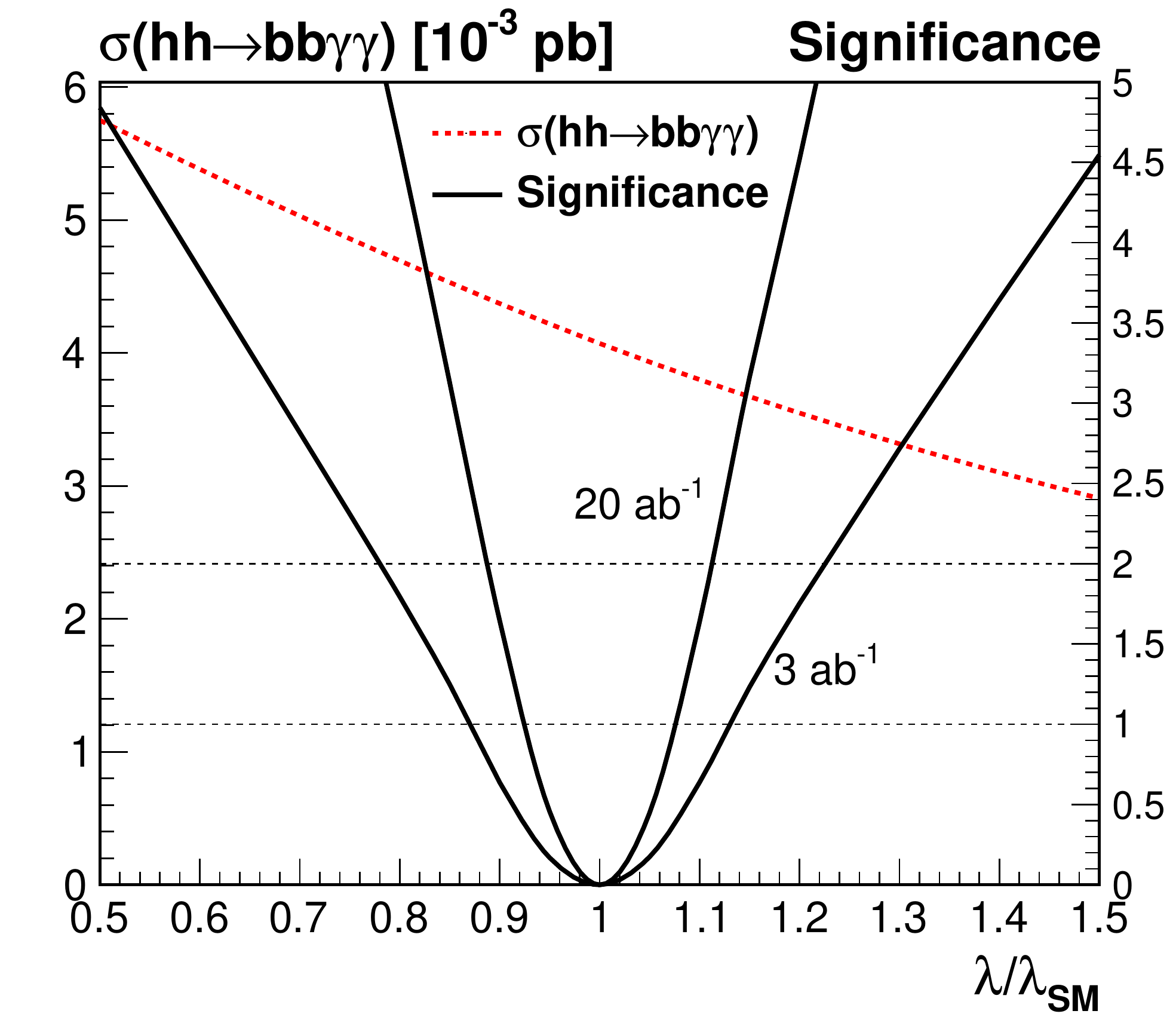}
  \caption{Signal cross section (red) and maximum significance (black) for observing
    an anomalous Higgs self-coupling at a 100~TeV hadron collider with
    an integrated luminosity of $3~\iab$ and $20~\iab$. The setup
    corresponds to Fig.~\ref{fig:total} for the high-luminosity LHC.}
\label{fig:total_nimatron}
\end{figure}

We can use exactly the same setup as in the main body of the paper to
estimate the impact of the signal kinematics at a future 100~TeV
collider with an integrated luminosity of $20~\iab$. Several studies,
largely based on a rate measurement with some background rejection
cuts, have shown that Higgs pair production can be probed at such a
machine with high precision~\cite{Contino:2016spe,nimatron_hh}.

\begin{figure}[t]
  \includegraphics[width=0.30\textwidth]{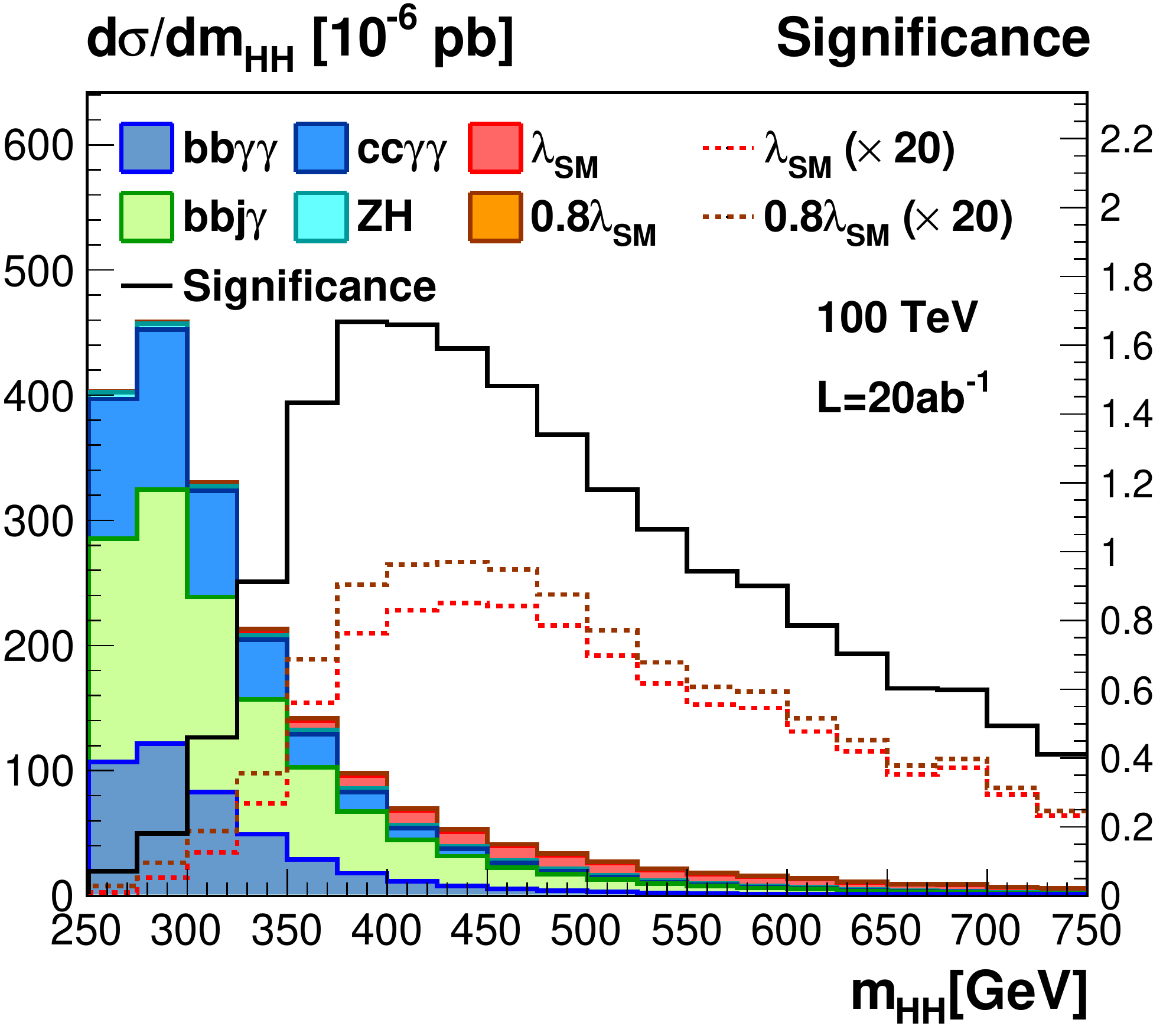}
  \hspace*{0.02\textwidth}
  \includegraphics[width=0.30\textwidth]{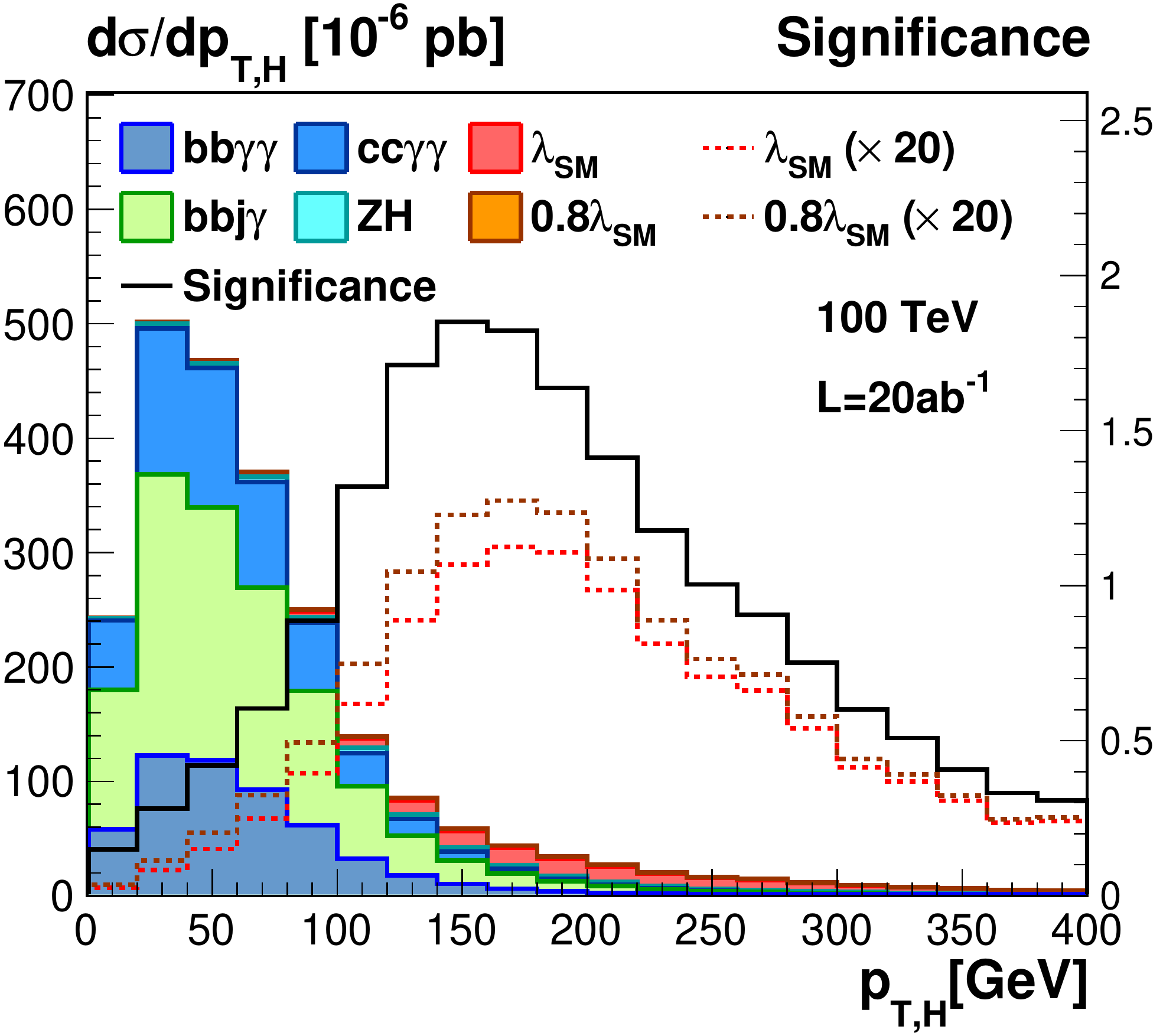}
  \hspace*{0.02\textwidth}
  \includegraphics[width=0.30\textwidth]{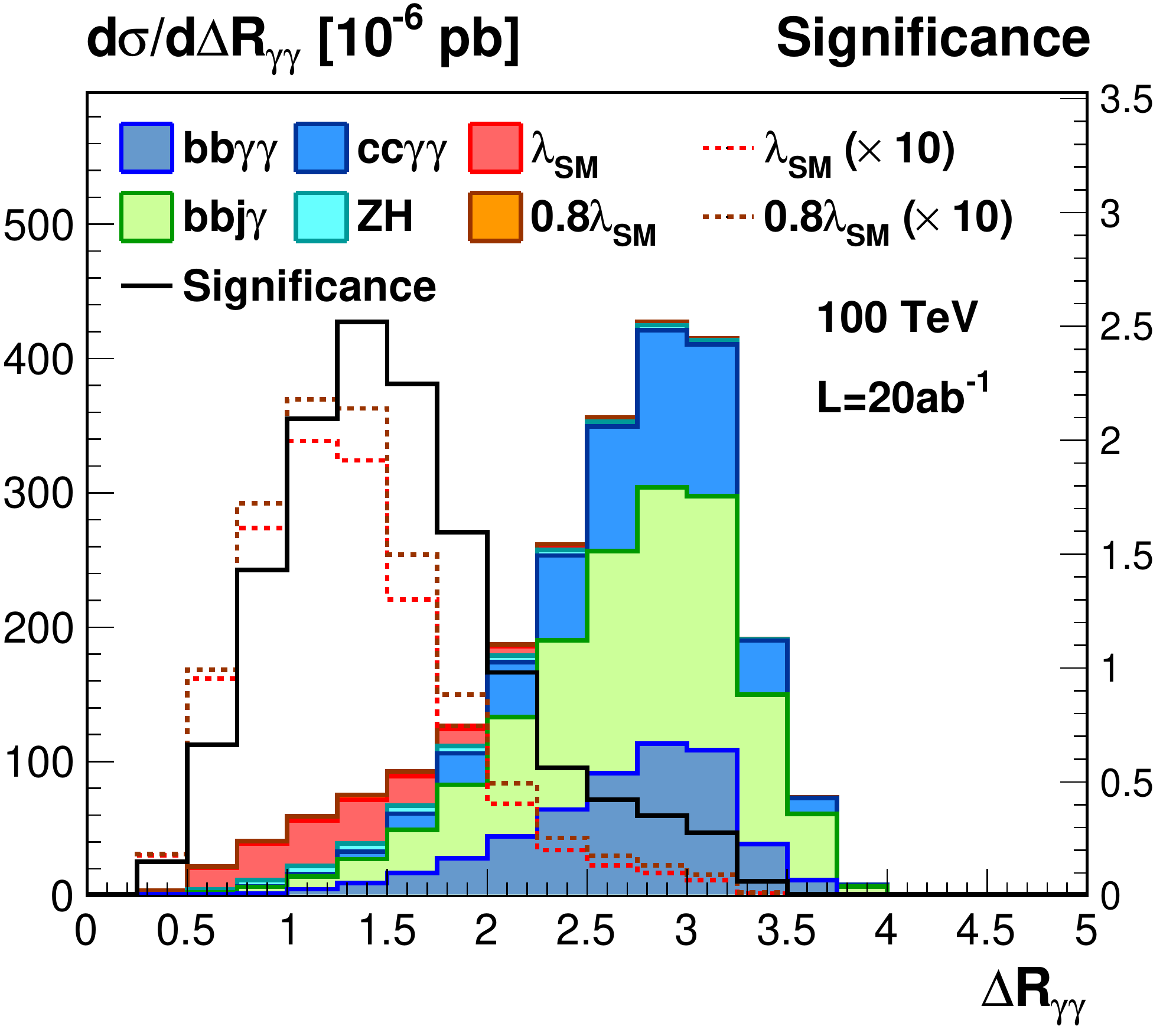} \\
  \includegraphics[width=0.30\textwidth]{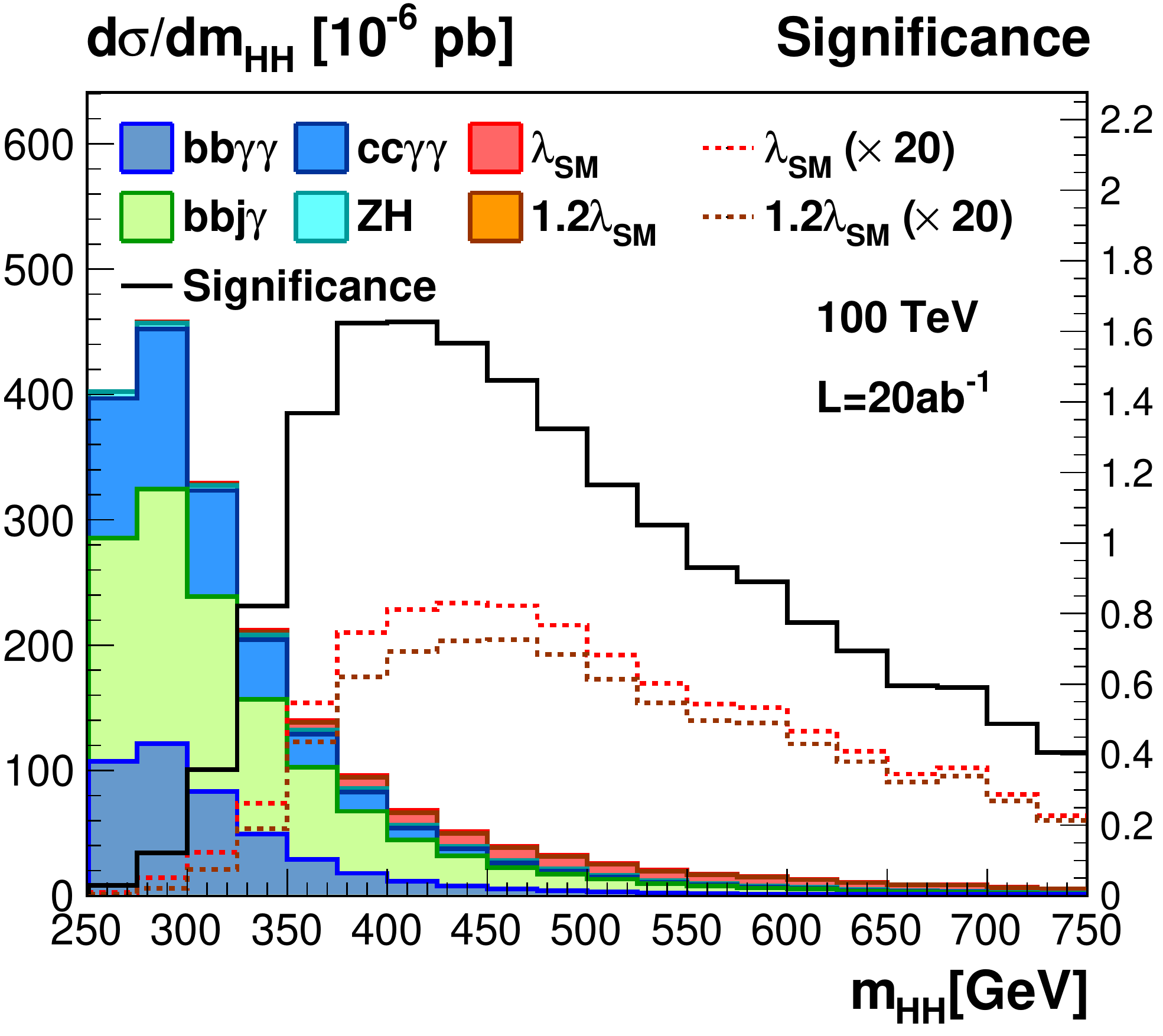}
  \hspace*{0.02\textwidth}
  \includegraphics[width=0.30\textwidth]{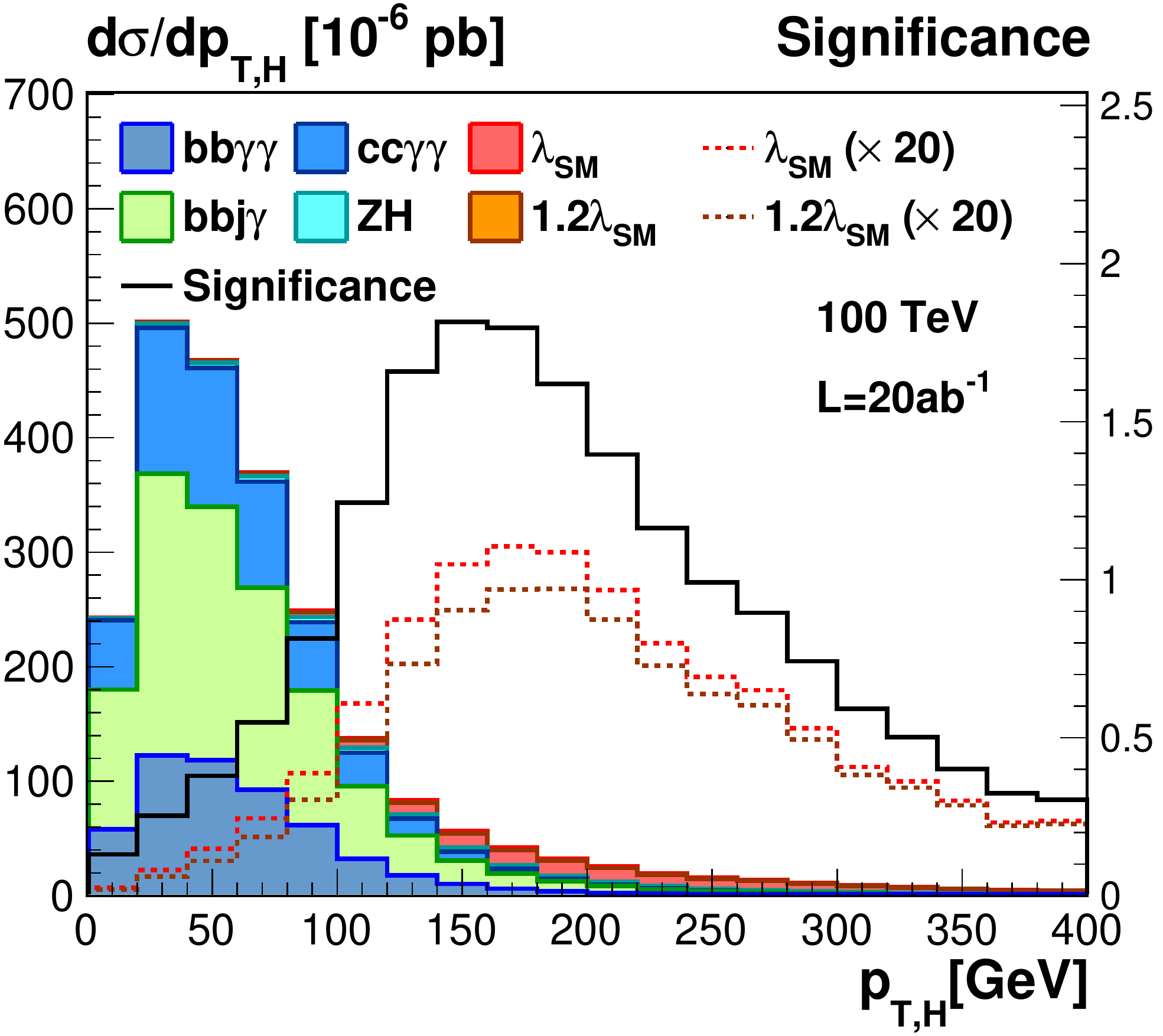}
  \hspace*{0.02\textwidth}
  \includegraphics[width=0.30\textwidth]{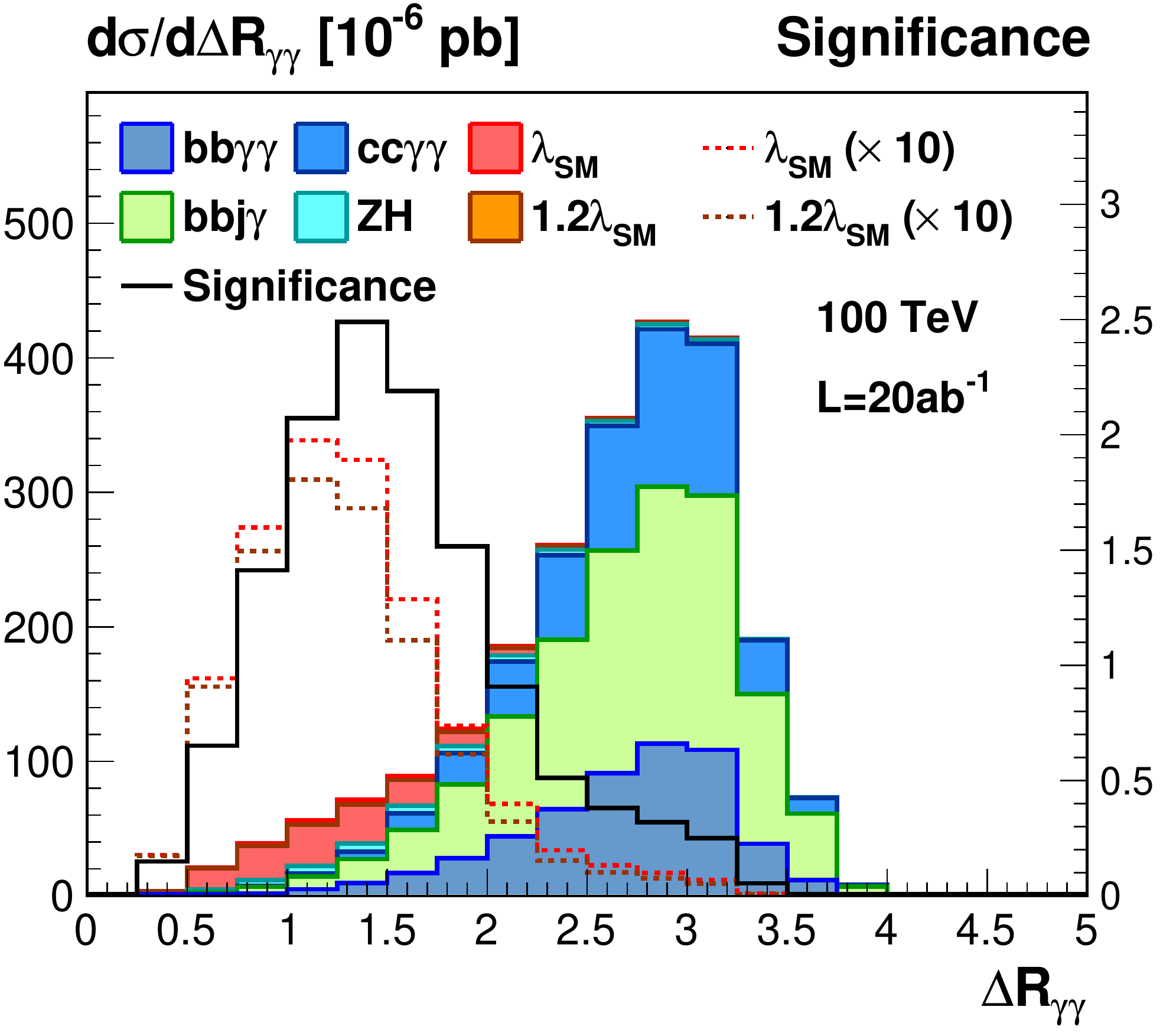} \\
  \vspace*{-0mm}
  \caption{Differential distributions assuming a modified Higgs
    self-coupling of $\lambda/\lambda_\text{SM}=0.8,1.2$ in the
    presence of backgrounds. The setup corresponds to
    Fig.~\ref{fig:coup_sb} for the high-luminosity LHC.}
\label{fig:coup_sb_nimatron}
\end{figure}

In Fig.~\ref{fig:total_nimatron} we show the maximum significance for
distinguishing a modified Higgs self-coupling from the Standard Model
at a future 100~TeV collider. The setup is exactly the same as for our
LHC analysis leading to the results shown in
Fig.~\ref{fig:coup_sb_nimatron}. This includes our detector smearing as well as
the trigger cuts. The signal cross section is now taken from
\textsc{MadMax} without an external normalization.  Relying on a
multi-variate analysis we can at best constrain the Higgs self-coupling
range to be
\begin{align}
\frac{\lambda}{\lambda_\text{SM}} = 0.92~...~1.07 \quad \text{at 68\% CL and for} \; 20~\iab 
\end{align}
and rule out
\begin{align}
\frac{\lambda}{\lambda_\text{SM}} = 0.89~...~1.11 \quad \text{at 95\% CL and for} \; 20~\iab \; . 
\end{align}
In Fig.~\ref{fig:coup_sb_nimatron} we also show the
significance distribution over phase space. It is essentially
identical to the 14~TeV case, because the relevant energy scale is
given by the Standard Model Higgs mass in the two propagators. As for
the LHC this implies that systematic and theoretical uncertainties
should not pose a major issue for this precision measurement, which is
driven by a very large number of signal events.\bigskip

These 100~TeV results quoted above appear, at first, to disagree with
some results shown in the literature. The main reason for this are
diverging assumptions about tagging efficiencies.  Since the Higgs
decay products are mostly concentrated in the central detector, we do
not expect the forward coverage of 100~TeV collider to significantly
affect our results. However, we find that assumptions about
$b$-tagging are crucial. The signal's $p_{T,b}$-distribution peaks
around $m_H/2$ and therefore in a regime with suppressed tagging
efficiency, about 45\% according to the parameterization in
App.~\ref{app:sim}.  An improved $b$-tagging efficiency at low
transverse momentum will significantly enhance the signal rate and
therefore improve the sensitivity for the triple-Higgs coupling. This
explains the better reach quoted in in Ref.~\cite{Contino:2016spe},
which assumes a constant 75\% tagging efficiency.

\section{Tagging efficiencies}
\label{app:sim}

Because the tagging and identification efficiencies have a crucial
effect on our results, we give the necessary details in this Appendix.
The $b$-tagging efficiency and the corresponding mistag rate depends
on the parton forming the tagged jet, the barrel respectively end-cap
position in the detector, and the transverse momentum. The
$b$-related tagging efficiencies we use in our analysis are
\begin{align}
  \text{$b$-jets} \qquad   & \epsilon_b = \Bigg\{
  \begin{array}{ll} 
    0.7\ \text{tanh}\left(0.01317 \; p_T/\text{GeV} - 0.062\right)  &\qquad |\eta|<1.2\\
    0.6\ \text{tanh}\left(0.01050 \; p_T/\text{GeV}  - 0.101\right)  &\qquad 1.2 <|\eta|<2.5 \\
  \end{array} \notag \\
  \text{$c$-jets} \qquad & \epsilon_c = \Bigg\{
  \begin{array}{ll} 
    0.1873\ \text{tanh}\left(0.01830 \; p_T/\text{GeV}  - 0.2196\right)  &\qquad |\eta|<1.2\\
    0.1898\ \text{tanh}\left(0.00997 \; p_T/\text{GeV}  - 0.1430\right)  &\qquad 1.2 <|\eta|<2.5 \\
  \end{array} \notag \\
  \text{light-flavor jets}  \qquad & \epsilon_j = 0.001 \; .
  \label{eq:b-tagging}
\end{align}
%
For the photons we follow Ref.~\cite{Butler:2020886} (Fig.~9.22),
which suggests a photon identification efficiency
$\epsilon^\prime_\gamma = 0.85$ and a $p_T$-dependent mistag rate
between $\epsilon^\prime_j=0.002$ and $0.0001$ for jets between 50~GeV
and 100~GeV:
\begin{align}
\text{photon} \qquad & \epsilon^\prime_\gamma = 0.85 \notag \\
\text{light-flavor jets} \qquad & \epsilon^\prime_j = \Bigg\{
  \begin{array}{lc} 
   0.01133\  \exp\left(-0.038 \; p_T/\text{GeV} \right) & \qquad p_T<100~\text{GeV}\\
   0.00025  & \qquad p_T>100~\text{GeV}.
  \end{array}
  \label{eq:gamma-tagging}
\end{align}


\end{document}